\documentclass[aps,prd,amsfonts,nofootinbib,preprintnumbers,nobalancelastpage]{revtex4-2}

\usepackage{graphicx}
\usepackage{mathtools}
\usepackage{epsfig}
\usepackage{xcolor}
\usepackage{ragged2e}
\usepackage{rotating}
\usepackage{amsmath}
\usepackage{braket}
\usepackage{caption}
\usepackage{subcaption}
\usepackage{cancel}
\usepackage{dsfont}
\usepackage{amssymb}
\usepackage{hyperref}
\usepackage{cleveref}
\usepackage{multirow}      
\usepackage{booktabs}      
\usepackage{array}         

\hypersetup{
  colorlinks=true,
  urlcolor=purple,
  anchorcolor=blue,
  citecolor=blue,
  filecolor=blue,
  linkcolor=purple,
  menucolor=blue,
  linktocpage=true
}

\begin{document}

\begin{flushleft}
\color{purple}{IFT-UAM/CSIC-26-160}
\end{flushleft}


\title{Unitarity and Positivity Constraints on C- or P-odd Anomalous Quartic Couplings}

\author{O.\ J.\ P.\ \'Eboli}
\email{eboli@if.usp.br}
\affiliation{Instituto de F\'{\i}sica,
             Universidade de S\~ao Paulo, S\~ao Paulo -- SP, Brazil.}

\author{Á.\ Lozano-Onrubia}
\email{alvaro.lozano.onrubia@csic.es}
\affiliation{Instituto de F\'{\i}sica Te\'{o}rica, IFT-UAM/CSIC, Cantoblanco, 28049, Madrid, Spain.}
\affiliation{Departamento de F\'{\i}sica Te\'{o}rica, Universidad Aut\'{o}noma de
  Madrid, Cantoblanco, 28049, Madrid, Spain.} 

\author{S.\ Rosende Herrero}
\email{samuel.rosende@uam.es}
\affiliation{Instituto de F\'{\i}sica Te\'{o}rica, IFT-UAM/CSIC, Cantoblanco, 28049, Madrid, Spain.}
\affiliation{Departamento de F\'{\i}sica Te\'{o}rica, Universidad Aut\'{o}noma de
  Madrid, Cantoblanco, 28049, Madrid, Spain.}

\begin{abstract}
  
In this work, we provide a comprehensive study of partial-wave unitarity and positivity constraints in the dimension-eight genuine anomalous quartic gauge coupling (aQGC) sector of the Standard Model Effective Field Theory (SMEFT). We extend previous partial-wave unitarity analyses to include all C- or P-odd operators, thereby completing the coverage of the full basis of 34 operators. We also derive an analytical set of necessary forward-limit positivity bounds that include all CP-odd Wilson coefficients. We demonstrate that non-vanishing CP-odd Wilson coefficients require non-vanishing contributions from the CP-even sector to satisfy the derived positivity conditions. Combining the positivity conditions with partial-wave unitarity, we obtain stronger bounds on twenty-three of the thirty-four Wilson coefficients in this sector of the SMEFT.

\end{abstract}


 \maketitle
\renewcommand{\baselinestretch}{1.15}
%

\section{Introduction}

Deviations from the Standard Model (SM) predictions in the high-energy tails of kinematic distributions could provide compelling evidence of new physics (NP) beyond the direct production reach of the CERN Large Hadron Collider (LHC)~\cite{ATLAS:2026fyh, VandenBossche:2025nvi, Franceschini:2017xkh, Farina:2016rws, ATLAS:2026wew, Alioli:2017jdo, Allwicher:2022gkm, Araz:2020zyh, Eboli:2006wa, Dedes:2020xmo, Sekulla:2016yku, Gillies:2026tpc, Hwang:2023wad, Degrande:2023iob}. Such searches often rely on the framework of effective field theory (EFT), which provides powerful tools to parametrize effects of heavy new physics beyond the SM at experimentally accessible energies without committing to a specific high-energy completion of the theory.\smallskip

A widely used framework for describing such effects is the Standard
Model Effective Field Theory (SMEFT). The SMEFT is built on the
premises that (i) the SM fields are the only relevant degrees of
freedom at the electroweak (EW) scale that is well below the
characteristic scale \(\Lambda\) of new physics; (ii) that the gauge
group of the effective description matches that of the SM; and,
furthermore, (iii) that the $SU(2)_L \times U(1)_Y$ gauge symmetry is realized linearly, {\it i.e.} the observed Higgs scalar belongs to an EW doublet.  Under these assumptions, the SMEFT Lagrangian can be organized as an expansion in operators of increasing canonical mass dimension, suppressed by inverse powers of the ultraviolet (UV) scale $\Lambda$, and it can be written as~\cite{Buchmuller:1985jz, Leung:1984ni, DeRujula:1991ufe, Hagiwara:1993ck, Gonzalez-Garcia:1999ije, Grzadkowski:2010es, Passarino:2012cb}
\begin{equation}
{\cal L}_{\rm SMEFT} = {\cal L}_{\text{SM}} + \sum_{n=5}^\infty\sum_i
\frac{f^{(n)}_i}{\Lambda^{n-4}} {\cal O}^{(n)}_i \;\; ,
\label{l:eff}
\end{equation}
where $\mathcal{L}_{\textrm{SM}}$ is the renormalizable SM Lagrangian, $f^{(n)}_i$ denotes dimensionless Wilson coefficients and the dimension-$n$ operators ${\cal O}^{(n)}_i$ comprise gauge-invariant combinations of SM fields, field strengths and covariant
derivatives.\smallskip

The EW gauge structure of the SM is one of its cornerstones. This
structure fixes the form of the gauge-boson self-interactions,
establishing correlations between the triple and quartic gauge
couplings. Deviations from the SM predictions for these interactions
could, therefore, provide evidence of NP, making this sector a
promising avenue for indirect searches. In this work, we focus on
deviations from the SM predictions for quartic gauge couplings
(QGCs). In SMEFT, anomalous contributions to QGCs first arise at
dimension six~\cite{Hagiwara:1993ck, Grzadkowski:2010es}. However,
these contributions are correlated with anomalous triple gauge
couplings (aTGCs) arising at the same order that are already well
constrained by EW gauge boson pair production analyses~\cite{ATLAS:2026fyh, CMS:2021cxr}. The present work is
instead interested in the class of \textit{genuine} anomalous QGCs
(aQGCs) without aTGC counterparts, which first appear at dimension
eight in SMEFT.\smallskip

General principles of local quantum field theories lead to constraints on low-energy EFTs. As is well known, NP entering through aTGCs and aQGCs can lead to the rapid growth of scattering
amplitudes~\cite{Cornwall:1974km}. At sufficiently high energies, this growth may give rise to violation of partial-wave unitarity, which calls into question the validity of the EFT and its perturbative description.  Furthermore, the assumption that the high-energy UV completion satisfies unitarity, causality, analyticity, and Lorentz invariance~\cite{Adams:2006sv, Vecchi:2007na, Bellazzini:2014waa, deRham:2017avq, Bi:2019phv, Remmen:2019cyz, Chala:2023xjy} leads to further constraints on the low-energy Wilson coefficients; these limits are known as positivity bounds.  Together, perturbative unitarity and positivity constrain both the range of perturbative validity of the EFT and the Wilson-coefficient combinations compatible with the assumed properties of its UV completion. However, the CP-odd sector of aQGCs has so far evaded a comprehensive systematic study in this respect.\smallskip

In this work, we precisely address this gap in the literature. First, we complete the partial-wave unitarity analysis of the dimension-eight aQGCs in Ref.~\cite{Almeida:2020ylr}, which focused on the twenty C-even and P-even operators. To this end, we derive the partial-wave unitarity bounds of the two additional C- and P-odd operators absent from the previous analysis, and subsequently extend the analysis to the twelve CP-odd operators that complete the basis~\cite{Durieux_aQGC}. In a second step, we derive for the first time a set of analytic necessary forward-limit positivity constraints encompassing all twelve CP-odd Wilson coefficients, extending previous studies of aQGC positivity~\cite{Zhang:2018shp, Remmen:2019cyz, Bi:2019phv,
Yamashita:2020gtt, Chang:2026ztn} and earlier analyses of
CP-odd correlations in the SMEFT~\cite{Remmen:2019cyz}.
We thus demonstrate that the existence of nonzero CP-odd Wilson coefficients requires a non-vanishing CP-even sector to satisfy the
positivity bounds (cf.\ Ref.~\cite{Remmen:2019cyz}). Finally, we combine the positivity constraints with partial-wave unitarity to obtain tighter bounds on twenty-three of the thirty-four Wilson coefficients.\smallskip

This paper is organized as follows. In Sec.~\ref{sec:forma}, we
present the aQGC operators considered in our analyses and describe the theoretical framework used to derive the partial-wave unitarity and forward-limit positivity constraints. Sec.~\ref{sec:results} contains the main results of the work, which are discussed in
Sec.~\ref{sec:discussion}. Several appendices contain additional
information on our analyses.\smallskip

\section{Analysis framework}
\label{sec:forma}

In the following, we first present the aQGC operators of interest for this work. We then briefly introduce the perturbative partial-wave unitarity and forward-limit positivity frameworks employed for their study. \smallskip

\subsection{Dimension-eight SMEFT aQGCs}
\label{sec:oper}

At mass dimension eight, the genuine aQGC sector of the SMEFT contains 22 independent CP-even~\cite{Eboli:2006wa, Eboli:2016kko} and 12 CP-odd operators~\cite{Remmen:2019cyz, Murphy:2020rsh}. The
construction of a complete and non-redundant basis was completed in
the recent Ref.~\cite{Durieux:2024zrg}, which also discusses their
individual properties under C and P transformations. We henceforth adopt the QGC basis and notation of Ref.~\cite{Durieux:2024zrg}. We refer the reader to this reference for further details. \smallskip

The set of CP-even aQGC operators can be categorized by field content into three distinct classes. In the $\Phi^4$ class we encounter three C- and P-even operators
\begin{equation}
\begin{array}{lcl}
{{\cal O}^S_{0}} =
  [D_{\mu}\Phi^{\dagger}D_{\nu}\Phi][D^{\mu}\Phi^{\dagger}D^{\nu}\Phi] \;,\\[2mm]
{{\cal O}^S_{1}} =
  [D^{\mu}\Phi^{\dagger}D_{\mu}\Phi][D^{\nu}\Phi^{\dagger}D_{\nu}\Phi] \;,\\[2mm]
{{\cal O}^S_{2}} =
  [D_{\mu}\Phi^{\dagger}D_{\nu}\Phi][D^{\nu}\Phi^{\dagger}D^{\mu}\Phi] \;,
\end{array}
\label{eq:cp_even_s}
\end{equation}
where $\Phi$ stands for the Higgs doublet and its covariant derivative is given by $D_{\mu}\Phi = (\partial_{\mu} + igW^j_{\mu} \tau^j +i\frac{1}{2}g'B_{\mu})\Phi$. Here, $B_{\mu}$ is the $U(1)_Y$ gauge field, $W_{\mu}^j$ the $j$-th component of the $SU(2)_L$ gauge field, and $\tau^j=\frac{\sigma^j}{2}$, where $\sigma^j$ is the $j$-th Pauli matrix. These are the only operators in the basis containing four Higgs doublets. This is the only class with uniform assignments under the individual actions of C and P transformations.\smallskip

Next, we display the nine CP-even aQGC operators included in the \textit{mixed} $\Phi^2 V^2$ class,
\begin{equation}
\begin{array}{lcl}
 {{\cal O}^M_{0}} =\frac{1}{2}
  [D^{\mu}\Phi^{\dagger}D_{\mu}\Phi]W^I_{\nu\rho}W^{I\nu\rho} \;,
&& 
 {{\cal O}^M_{1}}
=
   -\frac{1}{2}[D^{\mu}\Phi^{\dagger}D_{\nu}\Phi]W^I_{\mu\rho}W^{I\rho}_{\nu} \;,
\\[2mm]
 {{\cal O}^M_{2}} =
  [D^{\mu}\Phi^{\dagger}D_{\mu}\Phi]B_{\nu\rho}B^{\nu\rho} \;,
&& 
 {{\cal O}^M_{3}}
= -[D^{\mu}\Phi^{\dagger}D_{\nu}\Phi]B_{\mu\rho}B^{\rho}_{\nu} \;,
\\[2mm]
 {{\cal O}^M_{4}} = [D^{\mu}\Phi^{\dagger}\tau^I
  D_{\mu}\Phi]B^{\nu\rho}W^I_{\nu\rho} \;,
&& 
{{\cal O}^M_{5}} = [D^{\mu}\Phi^{\dagger}\tau^I
   D^{\nu}\Phi](B_{\mu}^{\rho}W^I_{\nu\rho}+B_{\nu}^{\rho}W^I_{\mu\rho}) \;,
\\[2mm]
 {{\cal O}^M_{7}} = [D^{\mu}\Phi^{\dagger}\tau^I\tau^J
  D^{\nu}\Phi]W^J_{\mu\rho}W^{I\rho}_{\nu} \;,
&& 
{{\cal O}^M_{8}} = i[D^{\mu}\Phi^{\dagger}\tau^I
   D^{\nu}\Phi](B_{\mu}^{\rho}\widetilde{W}^I_{\nu\rho}-B_{\nu}^{\rho}\widetilde{W}^I_{\mu\rho}) \;,
\\[2mm]
{{\cal O}^M_{9}} = \epsilon^{IJK}[D^{\mu}\Phi^{\dagger}\tau^I D^{\nu}\Phi](W^J_{\mu\rho}\widetilde{W}_{\nu}^{K\rho}-\widetilde{W}^J_{\mu\rho}W_{\nu}^{K\rho}) \, ,
\end{array}
\label{eq:cp_even_m}
\end{equation}
where $B_{\mu\nu}$ and $W_{\mu\nu}\equiv W^j_{\mu\nu}\tau^j$ are the
$U(1)_Y$ and $SU(2)_L$ field strengths, respectively. The dual field
strengths are defined as $\widetilde{B}^{\mu\nu}=\epsilon^{\mu\nu\rho\sigma}B_{\rho\sigma}/2$
and $\widetilde{W}^{I\mu\nu}=\epsilon^{\mu\nu\rho\sigma}
W_{\rho\sigma}^{I} /2$. We point out that operators
$\{{\cal O}^M_{0}, {\cal O}^M_{1}, {\cal O}^M_{2}, {\cal O}^M_{3},
{\cal O}^M_{4}, {\cal O}^M_{5}, {\cal O}^M_{7}\}$ are C- and P-even,
whereas $\{{\cal O}^M_{8},{\cal O}^M_{9}\}$ are C- and P-odd.\smallskip

The ten remaining CP-even operators, which are also C- and P-even,
belong to the \textit{tensorial} class $V^4$ given by
\begin{equation}
\begin{array}{lcl}
 {{\cal O}^T_{0}} =
  \frac{1}{4}W^I_{\mu\nu}W^{I\mu\nu}W^J_{\rho\sigma}W^{J\rho\sigma} \;,
&& 
 {{\cal O}^T_{1}}
=   \frac{1}{4}W^I_{\mu\nu}W^{J\mu\nu}W^I_{\rho\sigma}W^{J\rho\sigma} \;,
\\[2mm]
 {{\cal O}^T_{2}} =
  \frac{1}{4}W^I_{\mu\nu}W^{I\nu\alpha}W^J_{\alpha\beta}W^{J\beta\mu} \;,
&& 
 {{\cal O}^T_{3}}
=
   \frac{1}{4}W^I_{\mu\nu}W^{J\nu\alpha}W^I_{\alpha\beta}W^{J\beta\mu} \;,
\\[2mm]
 {{\cal O}^T_{4}} =
  \frac{1}{2}W^I_{\mu\nu}B^{\nu\alpha}W^I_{\alpha\beta}B^{\beta\mu} \;,
&& 
 {{\cal O}^T_{5}}
=   \frac{1}{2}B_{\mu\nu}B^{\mu\nu}W^I_{\rho\sigma}W^{I\rho\sigma} \;,
\\[2mm]
 {{\cal O}^T_{6}} =
  \frac{1}{2}B_{\mu\nu}W^{I\mu\nu}B_{\rho\sigma}W^{I\rho\sigma} \;,
&& 
 {{\cal O}^T_{7}}
=   \frac{1}{2}W^I_{\mu\nu}W^{I\nu\alpha}B_{\alpha\beta}B^{\beta\mu} \;,
\\[2mm]
 {{\cal O}^T_{8}} = B_{\mu\nu}B^{\mu\nu}B_{\rho\sigma}B^{\rho\sigma} \;,
&& 
 {{\cal O}^T_{9}}
=   B_{\mu\nu}B^{\nu\alpha}B_{\alpha\beta}B^{\beta\mu} \, .
\end{array}
\label{eq:cp_even_t}
\end{equation}

The CP-odd aQGC operators are exclusively contained within the mixed and tensorial classes. Within the former, we find the six operators
\begin{equation}
\begin{array}{lcl}
 {\cancel{\cal O}^M_{1}} = [D^{\mu}\Phi^{\dagger}D_{\mu}\Phi]B_{\nu\rho}\widetilde{B}^{\nu\rho},
&& 
 {\cancel{\cal O}^M_{2}}
=   [D^{\mu}\Phi^{\dagger}\tau^{I}D_{\mu}\Phi]B_{\nu\rho}\widetilde{W}^{I\,\nu\rho},
\\[2mm]
 {\cancel{\cal O}^M_{3}} = i[D^{\mu}\Phi^{\dagger}\tau^{I}D^{\nu}\Phi](B_{\mu\rho}W_{\nu}^{I\;\rho}-B_{\nu\rho}W_{\mu}^{I\;\rho}),
&&
 {\cancel{\cal O}^M_{4}} = [D^{\mu}\Phi^{\dagger}\tau^{I}D^{\nu}\Phi](B_{\mu\rho}\widetilde{W}_{\nu}^{I\;\rho}+B_{\nu\rho}\widetilde{W}_{\mu}^{I\;\rho}),
\\[2mm]
  {\cancel{\cal O}^M_{5}} =[D^{\mu}\Phi^{\dagger}D_{\mu}\Phi]W_{\nu\rho}^{I}\widetilde{W}^{I\,\nu\rho},
&&
  {\cancel{\cal O}^M_{6}}
   =i\,\epsilon^{IJK}[D^{\mu}\Phi^{\dagger}\tau^{I}D^{\nu}\Phi](W_{\mu\rho}^{J}\widetilde{W}_{\nu}^{K\;\rho}+\widetilde{W}_{\mu\rho}^{J}W_{\nu}^{K\;\rho})
    \;,

\end{array}
\label{eq:lind2}
\end{equation}
whereas the latter contains the remaining six purely tensorial operators
\begin{equation}
\begin{array} {lcl}
 {\cancel{\cal O}^T_{1}} =  B_{\mu\nu}B^{\mu\nu}B_{\rho\sigma}\widetilde{B}^{\rho\sigma},
&&
 {\cancel{\cal O}^T_{2}} = B_{\mu\nu}\widetilde{B}^{\mu\nu}W_{\rho\sigma}^{I}W^{I\,\rho\sigma},
\\[6pt]
 {\cancel{\cal O}^T_{3}} =  B_{\mu\nu}B^{\mu\nu}W_{\rho\sigma}^{I}\widetilde{W}^{I\,\rho\sigma},
&&
{\cancel{\cal O}^T_{4}}
= B_{\mu\nu}W^{I\,\mu\nu}B_{\rho\sigma}\widetilde{W}^{I\,\rho\sigma},
 \\[6pt]
{\cancel{\cal O}^T_{5}} =  W_{\mu\nu}^{I}W^{I\,\mu\nu}W_{\rho\sigma}^{J}\widetilde{W}^{J\,\rho\sigma},
 &&
 {\cancel{\cal O}^T_{6}} =
    W_{\mu\nu}^{I}W^{J\,\mu\nu}W_{\rho\sigma}^{I}\widetilde{W}^{J\,\rho\sigma} \;.
\end{array}
\label{eq:lan-t}
\end{equation}
Importantly, ${\cancel {\cal{O}}^M_3}$ is singled out as the only
operator exhibiting C-odd and P-even properties, whereas the remaining CP-odd operators are C-even and P-odd.\smallskip

On the whole, we shall consider the following dimension-eight
Lagrangian
${\cal L}_{\text{aQGC}} \subset {\cal
  L}_{\text{SMEFT}}^{\dim\textrm{-}8}$,
\begin{equation}
\label{eq:aQGCLag}
  {\cal L}_{\text{aQGC}} =
  \frac{1}{\Lambda^4}\sum_{j=0}^{2}
  f^S_j\,{\cal O}^S_j
  +
  \frac{1}{\Lambda^4}\sum_{\substack{j=0\\ j\neq 6}}^{9}
  f^M_j\,{\cal O}^M_j
  +
  \frac{1}{\Lambda^4}\sum_{j=0}^{9}
  f^T_j\,{\cal O}^T_j
  +
  \frac{1}{\Lambda^4}\sum_{j=1}^{6}
  \tilde f^M_j\,{\cancel{\cal O}}^M_j
  +
  \frac{1}{\Lambda^4}\sum_{j=1}^{6}
  \tilde f^T_j\,{\cancel{\cal O}}^T_j \, ,
\end{equation}
where the Wilson coefficients of the CP-even operators are ($f^S_j$,
$f^M_j$, $f^T_j$) while the CP-odd ones are ($\tilde f^M_j$,
$\tilde f^T_j$).\smallskip

Depending on the calculation at hand (cf.\ Secs.~\ref{sec:unit} and
\ref{sec:positivity}), pragmatism compels the use of the physical
basis over the electroweak basis tacitly assumed so far. In the
physical basis, we consider the fields after electroweak symmetry
breaking (EWSB) and in unitary gauge, where the vector bosons are
identified with the physical states $V=W^\pm,\, Z,\, A$, and $H$
denotes the observed Higgs boson. In this physical basis, the
operators in the mixed class generate quartic vertices of the form
$VVVV$, $VVVH$, and $VVHH$, while the tensorial class contributes only to pure quartic gauge vertices. Throughout this work, we consider the high-energy limit, where the Goldstone equivalence
theorem~\cite{Cornwall:1974km} relates amplitudes involving
longitudinal gauge bosons to the corresponding Goldstone-boson
amplitudes at leading order. \smallskip

\subsection{Partial-Wave Unitarity}
\label{sec:unit}

A simple dimensional analysis indicates that the scattering
amplitudes generated by the contact terms in Eq.~(\ref{eq:aQGCLag}) can grow as fast as $s^2/\Lambda^4$, where $\sqrt{s}$ denotes the center-of-mass energy. Therefore, the quantum-mechanical requirement that the $\mathcal{S}$-matrix be unitary can be used to constrain the behaviour of the amplitudes at high energies and thus
to bound the range of validity of the perturbative EFT description and limit its Wilson coefficients. In order to establish such unitarity bounds on the operators in Sec.~\ref{sec:oper}, we employ the partial-wave decomposition of the scattering amplitudes in the helicity basis for the processes at hand~\cite{Jacob:1959at}. Specifically, we consider $2\to2$ scattering processes $P_{1\lambda_1} P_{2\lambda_2} \rightarrow P_{3\lambda_3}P_{4\lambda_4}$, where $P_{i\lambda_i}$ ($i=1,2,3,4$)
denotes either an electroweak gauge boson with helicity $\lambda_i$ or the Higgs boson, for which $\lambda_i=0$. In the center-of-mass frame, the scattering amplitude of this process can be expanded in partial waves as
\begin{equation}
\mathcal{M}
({P_1}_{\lambda_1}{P_2 }_{\lambda_2} \to {P_3}_{\lambda_3}{P_4}_{\lambda_4})
=16 \pi \sum_{J=0}^{\infty}
\left (2 J+1 \right)~
\sqrt{1+\delta_{{P_1}_{\lambda_1}}^{{P_2}_{\lambda_2}}}
\sqrt{1+\delta_{{P_3}_{\lambda_3}}^{{P_4}_{\lambda_4}}}
d_{\lambda\mu}^{J}(\theta) ~e^{i M \varphi}
~ T^J({P_1}_{\lambda_1}{P_2 }_{\lambda_2} \to
{P_3}_{\lambda_3}{P_4}_{\lambda_4}) \, ,
\label{eq:helamp}
\end{equation}
where $J$ is the total angular momentum of the incoming state, $\theta$ and $\varphi$ are respectively the polar and azimuthal
scattering angles, and $d_{\lambda\mu}^{J}(\theta)$ is the Wigner rotation matrix in the convention of Ref.~\cite{Varshalovich:1988ifq}. We further define
$\lambda=\lambda_1-\lambda_2$, $\mu=\lambda_3-\lambda_4$, and
$M=\lambda-\mu$. \smallskip

In the high-energy limit where
$s \gg (M_{P_{1\lambda_1}}+M_{P_{2\lambda_2}})^2$, the elastic-channel unitarity reads
\begin{equation}
|T^J({P_1}_{\lambda_1}{P_2 }_{\lambda_2} \to {P_1}_{\lambda_1}{P_2}_{\lambda_2})| \le 1 \, .
\label{eq:unitcond}
\end{equation}
This condition can be generalized to the coupled-channel scenario~\cite{Lee:1977yc,Lee:1977eg}. In practice, for each value of $J$ one constructs the matrix $T^J$ in particle and helicity space and imposes Eq.~\eqref{eq:unitcond} on its eigenvalues. This
coupled-channel treatment is necessary in the presence of CP-odd
operators, whose contributions will appear in off-diagonal helicity or particle channels and, therefore, would be missed in a purely
elastic-channel analysis.\smallskip

A peculiarity of our analysis is that the contributions of the CP-odd operators to the partial-wave matrices $T^J$ are antisymmetric in the combined particle--helicity channel space. Furthermore, at leading order in perturbation theory, unitarity implies that these matrices are Hermitian. The combination of these two properties implies that the contributions of CP-odd operators to matrix elements are purely imaginary, with vanishing diagonal entries. In addition, the transformation properties under parity and time reversal also reduce the number of independent helicity amplitudes that need to be computed. These transformations are given by
\begin{align}
 T^J({P_1}_{\lambda_1}{P_2 }_{\lambda_2} \to {P_3}_{\lambda_3}{P_4}_{\lambda_4})
&=\eta_P(-1)^{\lambda_1-\lambda_2-\lambda_3+\lambda_4}
  T^J({P_1}_{-\lambda_1}{P_2 }_{-\lambda_2} \to {P_3}_{-\lambda_3}{P_4}_{-\lambda_4}),\label{parity}\\
 T^J({P_1}_{\lambda_1}{P_2 }_{\lambda_2} \to {P_3}_{\lambda_3}{P_4}_{\lambda_4})
&=\eta_T(-1)^{\lambda_1-\lambda_2-\lambda_3+\lambda_4} T^J({P_3}_{\lambda_3}{P_4 }_{\lambda_4} \to {P_1}_{\lambda_1}{P_2}_{\lambda_2}),
\label{time}
\end{align}
where $\eta_P=+1$ ($-1$) for P-even (P-odd) operators and $\eta_T = +1$ ($-1$) for CP-even (CP-odd) operators are the eigenvalues under parity and time-reversal.\smallskip

In our analysis, we evaluate the amplitude matrices $T^J$ corresponding to $J=0,1$ in particle and helicity space as functions
of the Wilson coefficients of the dimension-eight operators in
Eq.~(\ref{eq:aQGCLag}). Since by assumption we work in the high-energy limit, we shall only keep the leading-order $s^2$ contributions of the amplitudes corresponding to all combinations of incoming and outgoing states. It is convenient to work in the physical basis, where electric charge conservation allows us to further decompose the matrices into blocks classified by the charge $Q=0,1,2$ of the two-particle states involved. Altogether, we thus consider the blocks
\begin{equation}
\begin{array}{c | ccccccccccc r}
(Q,J)			&{\rm States}	&&&&&&&&&& \rm{Total} \\
\hline
\hline
(2,0) 		&	W^+_\pm W^+_\pm		&W^+_0W^+_0
&&&&&&&&&&3\\[+0.1cm]
\hline
(2,1) 		&W^+_\pm W^+_0 	&W^+_0W^+_\pm
&&&&&&&&&& 4	\\[+0.1cm]
\hline
(1,0) 		&	W^+_\pm Z_\pm		&W^+_0Z_0		
		&W^+_\pm \gamma_\pm		&W^+_0H		
&&&&&&&&6\\[+0.1cm]
\hline
(1,1) 		&	W^+_0Z_0		&W^+_\pm Z_0	&W^+_0Z_\pm
	&W^+_\pm Z_\pm	&	W^+_0\gamma_\pm	&W^+_\pm\gamma_\pm
		&W^+_0H					&W^+_\pm H	
&&&&14\\ [+0.1cm]
\hline
(0,0) 		&	W^+_\pm W^-_\pm		&W^+_0W^-_0		
		&Z_\pm Z_\pm	&Z_0Z_0	  	&	Z_\pm\gamma_\pm		
&\gamma_\pm\gamma_\pm	&Z_0H	&HH		
&&&&12\\[+0.1cm]
\hline
(0,1)		&	W^+_0W^-_0			&W^+_\pm W^-_0
		&W^+_0W^-_\pm	&W^+_\pm W^-_\pm	&	Z_\pm Z_0
		&Z_0Z_\pm 				&Z_\pm\gamma_\pm	&Z_0\gamma_\pm	
   &	Z_0H				&Z_\pm H	&\gamma_\pm H	
& 20\\[+0.1cm]
\end{array}
\label{eq:blocks}
\end{equation}
where upper and lower indices label charge and helicity, respectively, and the last column states the dimensionality of the
matrices. \smallskip

The scattering amplitudes were evaluated using FeynArts 3.9~\cite{Hahn:2000kx}, FeynCalc 10~\cite{Shtabovenko:2023idz} and FormCalc 9.9~\cite{Hahn:1998yk, Hahn:2000kx}, the latter of which uses FORM~\cite{Kuipers:2012rf} for symbolic manipulations. The model files were generated using FeynRules 2.0~\cite{Alloul:2013bka} with the implementation of
Refs.~\cite{Durieux_aQGC, Durieux:2024zrg}. In App.~\ref{app:tables}, we provide the independent amplitudes for all operators that are not simultaneously C- and P-even. For the complementary set, we refer the reader to the appendix of Ref.~\cite{Almeida:2020ylr}.\smallskip

To exemplify the process, we exhibit the ($Q=2$, $J=0$) matrix given
by
\begin{equation}
\frac{1}{96\pi}\frac{s^2}{\Lambda^4}
\left(
\begin{array}{@{}c@{\qquad}c@{\qquad}c@{}}
6f_1^T+3f_2^T+3f_3^T
& 0 &
\begin{gathered}
4f_0^T+8f_1^T+f_2^T+3f_3^T+i16(\tilde f_5^T+2\tilde f_6^T)
\end{gathered}
\\[2mm]
0 &
\mathclap{3f_0^S+f_1^S+f_2^S}
& 0
\\[2mm]
\begin{gathered}
4f_0^T+8f_1^T+f_2^T+3f_3^T-i16(\tilde f_5^T+2\tilde f_6^T)
\end{gathered}
& 0 &
6f_1^T+3f_2^T+3f_3^T
\end{array}
\right).
\label{eq:example}
\end{equation} 
The unitarity bounds as a function of $s$ are obtained by imposing Eq.~(\ref{eq:unitcond}) on the eigenvalues of this matrix. \smallskip

As a final remark, we point out that unitarity constraints can also be obtained from $2\to M$ amplitudes with $M>2$, which in some cases may exhibit a stronger growth in the high-energy limit than $2\to2$ amplitudes featuring the same operators~\cite{Bresciani:2025toe}. However, for the CP-even aQGCs the $2\to3$ unitarity bounds have been shown to be less stringent than the ones obtained from $2\to2$ scattering~\cite{Eboli:2026ccq}. For this reason, we restrict our analysis to $2\to2$ scattering amplitudes.\smallskip


\subsection{Forward-Limit Positivity}
\label{sec:positivity}

Positivity bounds provide constraints on Wilson coefficients that are complementary to the perturbative unitarity ones. Whereas partial-wave unitarity limits the size of scattering amplitudes at a given energy, positivity constrains the signs and relative sizes of the Wilson coefficients that can arise from a Lorentz-invariant, local, analytic and unitary UV completion. Under these assumptions, together with crossing symmetry and suitable high-energy boundedness, twice-subtracted forward dispersion relations imply that the coefficient of the $s^2$ term in the low-energy expansion of an appropriately pole-subtracted forward elastic amplitude is
non-negative \cite{Adams:2006sv, Vecchi:2007na, Bellazzini:2014waa, deRham:2017avq, Zhang:2018shp, Remmen:2019cyz, Bi:2019phv, Chala:2023xjy}. Consequently, once the relevant contributions to the
amplitude are consistently accounted for, Wilson-coefficient
configurations that violate these conditions cannot arise from a UV
completion satisfying the aforementioned assumptions. Therefore,
positivity provides a powerful consistency check that restricts the
allowed Wilson-coefficient parameter space of an EFT.  \smallskip

In this work, we focus on the tree-level dimension-eight contact
contributions to forward scattering amplitudes in the high-energy
limit. Denoting by $\mathcal X_A$ an orthonormal one-particle basis state, with $A$ collecting the particle --- and, where applicable, the helicity --- labels, we write, in the massless forward limit,
\begin{equation}
    \mathcal M
    \left(
        \mathcal X_A\mathcal X_B
        \to \mathcal X_C\mathcal X_D;s,t
    \right)\big|_{t=0}
    =
    \frac{s^2}{\Lambda^4}\,
    \mathcal C_{AB,CD}\,,
\label{eq:forward_C_definition}
\end{equation}
where $\mathcal C_{AB,CD}$ denotes the forward coefficient tensor
whose entries are linear combinations of the dimension-eight Wilson
coefficients. \smallskip

In the elastic approach, positivity is imposed on product states $|u\rangle\otimes|v\rangle$, where each incoming particle is an independent superposition of admissible one-particle states \cite{Zhang:2018shp,Bi:2019phv,Yamashita:2020gtt,Li:2021lpe}. We therefore consider
\begin{equation}
\begin{aligned}
    |u\rangle
    &=\sum_A u_A|\mathcal X_A\rangle \;,
    &\qquad
    u\in\mathbb C^{N_1}\;,
    &\qquad
    u^\dagger u=1 \;,
    \\
    |v\rangle
    &=\sum_B v_B|\mathcal X_B\rangle \;,
    &\qquad
    v\in\mathbb C^{N_2} \;,
    &\qquad
    v^\dagger v=1 \;,
\end{aligned}
\label{eq:positivity_states}
\end{equation}
where $N_1$ and $N_2$ denote the number of one-particle basis
states available to the first and second incoming particles,
respectively. We retain complex coefficients to account for
arbitrary relative phases between the states.\footnote{Ref.~\cite{Yamashita:2020gtt} shows that real superpositions are sufficient when working in a basis of real modes with manifest crossing symmetry. However, this simplification does not directly apply to the helicity basis used here, where helicity states are complex combinations of linearly polarized states. We therefore retain complex superpositions.}
The coefficient of the corresponding forward elastic amplitude
is then
\begin{equation}
    \mathcal C(u,v)
    =
    \sum_{A,B,C,D}
    u_C^*v_D^*\,
    \mathcal C_{AB,CD}\,
    u_Av_B\; .
\label{eq:Cuv_pos}
\end{equation}
A complete treatment of these forward-elastic bounds would require imposing $\mathcal C(u,v)\geq0$ for arbitrary $u$ and $v$. However, we do not attempt to solve this condition globally, nor to reconstruct the optimal positivity cone \cite{Yamashita:2020gtt,Li:2021lpe,Chang:2026ztn,Zhang:2020jyn}. Instead, we extract a finite set of necessary forward-positivity conditions by choosing representative directions for $u$. For each such choice, we define the Hermitian matrix
\begin{equation}
    K_{DB}(u)
    =
    \sum_{A,C}
    u_C^*\,
    \mathcal C_{AB,CD}\,
    u_A ,
\label{eq:Ku_definition}
\end{equation}
so that
\begin{equation}
    \mathcal C(u,v)
    =
    v^\dagger K(u)v \geq 0,
    \quad \forall\,v
    \qquad\Longleftrightarrow\qquad
    K(u)\succeq 0 .
\label{eq:Cuv_Ku_relation}
\end{equation}
Thus, for each selected direction $u$, we examine the eigenvalues of
$K(u)$ and retain the resulting positivity conditions that can be
expressed as linear or quadratic inequalities in the Wilson
coefficients. We complement these with constraints obtained from the
non-negativity of the diagonal entries and all $2\times2$ principal
minors of $K(u)$, as also considered in Ref.~\cite{Chakraborty:2024ciu}. After combining the inequalities
obtained through both procedures and removing repetitions and
established redundancies, we retain a finite set of necessary
positivity bounds up to quadratic order in the Wilson coefficients.
\smallskip

Although the calculation could be performed directly in the physical
basis, the electroweak basis makes the field-content structure of the amplitudes more transparent. We therefore use the basis
\begin{equation}
    \left\{
    B_-,B_+,
    W^1_-,W^1_+,
    W^2_-,W^2_+,
    W^3_-,W^3_+,
    \phi_1,\phi_2,\phi_3,H
    \right\} \;,
\label{eq:EW_pos_basis}
\end{equation}
where $\pm$ denote transverse helicities, $\phi_i$ are the Goldstone
modes and $H$ is the Higgs mode. Longitudinal gauge-boson amplitudes
are represented by the corresponding Goldstone amplitudes in the
high-energy approximation.\smallskip

The matrices are organized according to the electroweak field content of the operators. The purely transverse contributions of $B^4$, $W^4$ and $B^2W^2$ classes involve, respectively, the one-particle subspaces spanned by
\begin{equation}
    \{B_-,B_+\},\qquad
    \{W^I_-,W^I_+\},\qquad
    \{B_-,B_+,W^I_-,W^I_+\} \;,
\end{equation}
with $I=1,2,3$. The mixed contributions of the $\Phi^2B^2$,
$\Phi^2W^2$ and $\Phi^2BW$ classes similarly involve the subspaces
spanned, respectively, by
\begin{equation}
  \{B_-,B_+,\phi_1,\phi_2,\phi_3,H\}\;,\qquad
  \{W^I_-,W^I_+,\phi_1,\phi_2,\phi_3,H\}\;,\qquad\{B_-,B_+,W^I_-,W^I_+,\phi_1,\phi_2,\phi_3,H\}\;.
\end{equation}

We first analyze positivity sector by sector, restricting the Wilson
coefficients and the one-particle states to the corresponding active
subspaces. For the $B^4$, $W^4$, $\Phi^2B^2$, and $\Phi^2W^2$ sectors, this isolated analysis already yields nontrivial positivity bounds while allowing nonzero Wilson coefficients. By contrast, treating the $B^2W^2$ and $\Phi^2BW$ sectors in isolation can force certain coefficient combinations to vanish, motivating their combined analysis with the other sectors.\smallskip

The $\Phi^2BW$ sector does not contain diagonal elastic support by
itself. For a fixed scalar direction $u_\Phi$, the corresponding
gauge-boson block of $K(u_\Phi)$ has the schematic form
\begin{equation}
K_{\Phi^2BW}(u_\Phi)\sim
\begin{pmatrix}
0 & X\\
X^\dagger & 0
\end{pmatrix},
\end{equation}
where the block $X$ contains combinations of the $\Phi^2BW$ Wilson coefficients. Requiring $K(u_\Phi)\succeq0$ forces $X=0$. Therefore, generic nonzero $\Phi^2BW$ coefficients require the
diagonal support provided by the $\Phi^2B^2$ and $\Phi^2W^2$
sectors. We consequently consider the combined mixed-sector matrix
\begin{equation}
K_M(u)=K_{\Phi^2B^2}(u)+K_{\Phi^2W^2}(u)+K_{\Phi^2BW}(u)  \;.
\end{equation}

The $B^2W^2$ sector differs in that it already contains diagonal
elastic $BW\to BW$ amplitudes. Nevertheless, these amplitudes do not
necessarily provide the diagonal support required by every off-diagonal combination in the contracted matrices $K(u)$. By the
same positive-semidefiniteness argument as above, an isolated $B^2W^2$ analysis can therefore force particular Wilson-coefficient
combinations to vanish, while allowing other $BW$ amplitudes to remain nonzero.  Including the $B^4$ and $W^4$ contributions can provide the missing diagonal support, allowing such combinations to remain nonzero subject to correlated positivity inequalities. We therefore consider the combined transverse-sector matrix
\begin{equation}
    K_T(u)
    =
    K_{B^4}(u)
    +K_{W^4}(u)
    +K_{B^2W^2}(u)\;,
\label{eq:CT_pos}
\end{equation}
where the representative directions $u$ are chosen within the one-particle subspaces appropriate to each analysis. For the transverse sector, they include pure $B$ helicity directions, $W$ directions with helicity and weak-isospin superpositions, including a non-factorizable weak-isospin--helicity direction, and mixed $B$--$W$
directions. The mixed-sector analysis additionally includes scalar,
$B$--$\Phi$ and $W$--$\Phi$ directions\footnote{A more comprehensive analysis could include simultaneous $B$--$W$--$\Phi$ superpositions, potentially yielding additional positivity constraints. Nevertheless, the directions considered here already provide nontrivial bounds on all the CP-odd coefficients.}. The choices used to obtain the displayed bounds, including their relative phases, are specified in App.~\ref{app:uchoices}. \smallskip

Although a more general positivity analysis could include simultaneous contributions from the purely scalar, mixed, and transverse sectors, together with their correlations, we restrict our study to the tensorial and mixed sectors separately. Consequently, bounds derived from external-state superpositions after neglecting contributions from other sectors apply to the corresponding restricted coefficient space. A complete analysis including all three sectors and their correlations is beyond the scope of this work. \smallskip


\section{Results}
\label{sec:results}

We present our results for three benchmark scenarios. The first considers perturbative unitarity with one non-vanishing C- or P-odd
operator at a time. The second allows the simultaneous presence of CP-even and CP-odd operators, imposing only perturbative unitarity. Finally, the third benchmark supplements these constraints with forward-limit positivity bounds. The tabulated bounds and positivity inequalities are implemented in accompanying Python and Mathematica codes, available in the GitHub repository~\cite{aQGCchecker}.\smallskip

\subsection{First Benchmark: Unitarity Bounds On Isolated Operators
  Of The C- Or P-odd Sector}
\label{single_ops}

In this first benchmark, we begin by studying perturbative unitarity alone with only one active direction at a time in the space of C- or P-odd Wilson coefficients. This setup is fully consistent as an EFT benchmark, but without positivity it should not be interpreted as representative of a generic UV completion. Table~\ref{tab:ls_individual} presents the corresponding bounds of such coefficients for the mixed and tensorial operators. The strongest constraints are obtained from the partial-wave matrices ($Q=0$, $J=0$) for all operators, except ${\cal O}^M_8$, ${\cal O}^M_9$, ${\cancel{\cal O}^M_3}$, and ${\cancel{\cal O}^M_6}$. For these operators, the leading contribution arises in the ($Q=0$, $J=1$) channel, resulting in comparatively weaker bounds. Furthermore, the limits on the $C$- or $P$-odd tensorial operators are very tight with limits comparable to the corresponding ones for $C$- and $P$-even tensorial operators (cf.\ Ref.~\cite{Almeida:2020ylr}).\smallskip


\begin{table}

\makebox[\textwidth][c]{%
\begin{tabular}{|c|c|c||c|c|c|}
\hline
 & \multicolumn{2}{c||}{\rule{0pt}{2.5ex}Bounds on $f^M_i$ and $\tilde{f}^M_{i}$} 
 & 
 & \multicolumn{2}{c|}{\rule{0pt}{2.5ex}Bounds on $\tilde{f}^T_{i}$} 
\\ \cline{2-3}\cline{5-6}

\shortstack{Wilson\\coefficient}
& Analytical
& For $\sqrt{s}<1.5\, (3)$ TeV
& \shortstack{Wilson\\coefficient}
& Analytical
& For $\sqrt{s}<1.5\, (3)$ TeV
\\
\hline
&&&&&\\[-0.2cm]

$\left|\dfrac{f^M_{8}}{\Lambda^4}\right|$
& $\dfrac{512\sqrt{3}}{5} \, \pi \, s^{-2}$
& $110 \, (6.9)$ TeV$^{-4}$
& $//$
& $//$
& $//$
\\[+0.2cm]

$\left|\dfrac{f^M_{9}}{\Lambda^4}\right|$
& $48 \, \pi \, s^{-2}$
& $30 \, (1.9)$ TeV$^{-4}$
& $//$
& $//$
& $//$
\\[+0.2cm]

$\left|\dfrac{\tilde{f}^M_{1}}{\Lambda^4}\right|$
& $8\sqrt{2} \, \pi \, s^{-2}$
& 7.0 (0.44) TeV$^{-4}$
& $\left|\dfrac{\tilde{f}^T_{1}}{\Lambda^4}\right|$
& $\dfrac{12}{5} \, \pi \, s^{-2}$
& 1.5 (0.093) TeV$^{-4}$
\\[+0.2cm]

$\left|\dfrac{\tilde{f}^M_{2}}{\Lambda^4}\right|$
& $32 \, \pi \, s^{-2}$
& 20 (1.2) TeV$^{-4}$
& $\left|\dfrac{\tilde{f}^T_{2}}{\Lambda^4}\right|$
& $\dfrac{4\sqrt{3}}{3} \, \pi \, s^{-2}$
& 1.4 (0.090) TeV$^{-4}$
\\[+0.2cm]

$\left|\dfrac{\tilde{f}^M_{3}}{\Lambda^4}\right|$
& $\dfrac{512\sqrt{3}}{5} \, \pi \, s^{-2}$
& 110 (6.9) TeV$^{-4}$
& $\left|\dfrac{\tilde{f}^T_{3}}{\Lambda^4}\right|$
& $\dfrac{4\sqrt{3}}{3} \, \pi \, s^{-2}$
& 1.4 (0.090) TeV$^{-4}$
\\[+0.2cm]

$\left|\dfrac{\tilde{f}^M_{4}}{\Lambda^4}\right|$
& $64 \, \pi \, s^{-2}$
& 40 (2.5) TeV$^{-4}$
& $\left|\dfrac{\tilde{f}^T_{4}}{\Lambda^4}\right|$
& $6 \, \pi \, s^{-2}$
& 3.7 (0.23) TeV$^{-4}$
\\[+0.2cm]

$\left|\dfrac{\tilde{f}^M_{5}}{\Lambda^4}\right|$
& $\dfrac{8\sqrt{6}}{3}\, \pi \, s^{-2}$
& 4.1 (0.25) TeV$^{-4}$
& $\left|\dfrac{\tilde{f}^T_{5}}{\Lambda^4}\right|$
& $\dfrac{12}{11} \, \pi \, s^{-2}$
& 0.68 (0.042) TeV$^{-4}$
\\[+0.2cm]

$\left|\dfrac{\tilde{f}^M_{6}}{\Lambda^4}\right|$
& $\dfrac{384}{5} \, \pi \, s^{-2}$
& 48 (3.0) TeV$^{-4}$
& $\left|\dfrac{\tilde{f}^T_{6}}{\Lambda^4}\right|$
& $\dfrac{12}{7} \, \pi \, s^{-2}$
& 1.1 (0.066) TeV$^{-4}$
\\

\hline
\end{tabular}%
}

\vspace{2mm}

\refstepcounter{table}
\label{tab:ls_individual}

\makebox[\textwidth][c]{%
\begin{minipage}{0.95\textwidth}
\small
\justifying
\noindent\textbf{TABLE~\thetable.}
Perturbative partial-wave unitarity constraints on the Wilson coefficients $f^M_i$, $\tilde{f}^M_{i}$ and $\tilde{f}^T_{i}$ of the
corresponding C- or P-odd operators ${\cal O}^M_i$, $\cancel{\cal O}^M_i$ and $\cancel{\cal O}^T_i$ for the first benchmark in the limit of one active direction at a time.
\end{minipage}%
}

\label{tab:scenario_one}
\end{table}

\subsection{Second Benchmark: Unitarity Bounds On Simultaneously Active CP-even And CP-odd Coefficients}

In our second benchmark scenario, we obtain the partial-wave unitarity bound on a given Wilson coefficient marginalizing over the remaining 33 couplings. This scenario leads to the loosest constraints on the Wilson coefficients but it might require the existence of additional interactions to tame the growth of the scattering amplitudes. This analysis is simplified by the fact that the marginalization over just one operator sector leads to the same limits as marginalizing over the whole parameter space; for a proof of this fact see Appendix~\ref{app:block_proof}. Therefore, we just have to analyze each sector separately. \smallskip

To illustrate this property, we display the $Q=1$ block of the $J=0$ partial-wave matrix for tensorial operators, restricted to its four non-vanishing transverse channels. We denote by $T_{\mathrm{phys}}^{(1,0)}$ ($T_{\mathrm{EW}}^{(1,0)}$) the scattering matrix in the physical (electroweak) basis given by
\begin{align}
\mathcal B_{\mathrm{phys}}&=
(W^+_-Z_-,\,W^+_+Z_+,\,W^+_-\gamma_-,\,W^+_+\gamma_+) \; ,
\nonumber\\
\mathcal B_{\mathrm{EW}}&=
(W^+_-W^3_-,\,W^+_+W^3_+,\,W^+_-B_-,\,W^+_+B_+) \; .
\end{align}
Introducing the Hermitian matrices
\begin{equation}
\mathcal W=
\begin{pmatrix}
a_W & b_W\\
b_W^* & a_W
\end{pmatrix} \; ,
\qquad
\mathcal{BW}=
\begin{pmatrix}
a_{BW} & b_{BW}\\
b_{BW}^* & a_{BW}
\end{pmatrix} \; ,
\end{equation}
where
\begin{align}
a_W &=3(2f_1^T+f_2^T+f_3^T) \; ,\nonumber\\
b_W &=4f_0^T+8f_1^T+f_2^T+3f_3^T
+16i(\tilde f_5^T+2\tilde f_6^T) \; ,\nonumber\\
a_{BW} &=3(f_4^T+2f_6^T+f_7^T) \;,\nonumber\\
b_{BW} &=3f_4^T+4f_5^T+8f_6^T+f_7^T
+8i(\tilde f_2^T+\tilde f_3^T+2\tilde f_4^T) \; ,
\end{align}
the partial-wave matrix in the physical basis takes the form
\begin{equation}
T_{\mathrm{phys}}^{(1,0)}
=\frac{s^2}{96\pi\Lambda^4}
\begin{pmatrix}
c_W^2\mathcal W+s_W^2\mathcal{BW}
& s_Wc_W(\mathcal W-\mathcal{BW})\\
s_Wc_W(\mathcal W-\mathcal{BW})
& s_W^2\mathcal W+c_W^2\mathcal{BW}
\end{pmatrix} \, ,
\end{equation}
where $c_W=\cos\theta_W$ and $s_W=\sin\theta_W$ are the cosine and
sine of the weak mixing angle. The two bases are related by the orthogonal transformation
\begin{equation}
U_1=
\begin{pmatrix}
c_W\mathds{1}_2 & -s_W\mathds{1}_2\\
s_W\mathds{1}_2 & c_W\mathds{1}_2
\end{pmatrix} \; ,
\end{equation}
which yields
\begin{equation}\label{eq:unitarity_block_example}
T_{\mathrm{EW}}^{(1,0)}
=U_1^\dagger T_{\mathrm{phys}}^{(1,0)}U_1
=\frac{s^2}{96\pi\Lambda^4}
\begin{pmatrix}
\mathcal W & 0\\
0 & \mathcal{BW}
\end{pmatrix} \; .
\end{equation}
The two blocks depend exclusively on the $W^4$ and $B^2W^2$ coefficients, respectively. More generally, as shown in App.~\ref{app:block_proof}, the contributions of each sector can be
isolated through fixed projectors, allowing the marginalized unitarity bounds to be obtained from the individual sectors. \smallskip

We organize the analyses into six sectors, defined by both their mixed or tensorial operator structure (cf.\ Sec.~\ref{sec:positivity}) and the particle content of the corresponding scattering amplitudes. Specifically, we consider the tensorial sectors $B^4$, $W^4$ and $B^2W^2$ as well as the mixed sectors $\Phi^2B^2$, $\Phi^2W^2$ and $\Phi^2BW$. The resulting bounds are presented in Table~\ref{tab:multi}. Comparing Tables~\ref{tab:ls_individual} and~\ref{tab:multi}, we learn that marginalization relaxes the limits on the Wilson coefficients $\tilde f^M_{2,4,6}$ of the mixed class, while for the tensorial class all CP-odd bounds are loosened except that on $\tilde f^T_1$. \smallskip

It is important to keep in mind that each bound of this benchmark
corresponds to the maximum absolute value attainable by the coefficient under consideration, while the others are allowed to vary consistently with perturbative unitarity. These extrema generally occur at different points in parameter space and should not be interpreted as simultaneously attainable values.  Moreover, the bounds on the C- and P-even coefficients coincide with those obtained in Ref.~\cite{Almeida:2020ylr} for the complete mixed and tensorial systems, remaining unchanged upon the inclusion of C- or P-odd operators. We therefore refrain from explicitly presenting them here and defer a detailed discussion of the relevant eigenvalues and their properties to the following subsection.\smallskip

\begin{table}

\makebox[\textwidth][c]{%
\begin{tabular}{|c|c|c||c|c|c|}
\hline
 & \multicolumn{2}{c||}{\rule{0pt}{2.5ex}Bounds on $f^M_{i}$ and $\tilde{f}^M_{i}$} 
 & 
 & \multicolumn{2}{c|}{\rule{0pt}{2.5ex}Bounds on $\tilde{f}^T_{i}$} 
\\ \cline{2-3}\cline{5-6}

\shortstack{Wilson\\coefficient}
& Analytical
& For $\sqrt{s}<1.5\, (3)$ TeV
& \shortstack{Wilson\\coefficient}
& Analytical
& For $\sqrt{s}<1.5\, (3)$ TeV
\\
\hline
&&&&&\\[-0.2cm]

$\left|\dfrac{f^M_{8}}{\Lambda^4}\right|$
& $\dfrac{512\sqrt{3}}{5} \, \pi \, s^{-2}$
& $110 \, (6.9)$ TeV$^{-4}$
& $//$
& $//$
& $//$
\\[+0.2cm]

$\left|\dfrac{f^M_{9}}{\Lambda^4}\right|$
& $48 \, \pi \, s^{-2}$
& $30 \, (1.9)$ TeV$^{-4}$
& $//$
& $//$
& $//$
\\[+0.2cm]

$\left|\dfrac{\tilde{f}^M_{1}}{\Lambda^4}\right|$
& $8\sqrt{2} \, \pi \, s^{-2}$
& 7.0 (0.44) TeV$^{-4}$
& $\left|\dfrac{\tilde{f}^T_{1}}{\Lambda^4}\right|$
& $\dfrac{12}{5} \, \pi \, s^{-2}$
& 1.5 (0.093) TeV$^{-4}$
\\[+0.2cm]

$\left|\dfrac{\tilde{f}^M_{2}}{\Lambda^4}\right|$
& $4(5+8\sqrt{3})\, \pi \, s^{-2}$
&47 (2.9) TeV$^{-4}$
& $\left|\dfrac{\tilde{f}^T_{2}}{\Lambda^4}\right|$
& $\dfrac{2}{5}(3+4\sqrt{3}) \, \pi \, s^{-2}$
& 2.5 (0.15) TeV$^{-4}$
\\[+0.2cm]

$\left|\dfrac{\tilde{f}^M_{3}}{\Lambda^4}\right|$
& $\dfrac{512\sqrt{3}}{5}\, \pi \, s^{-2}$
& 110 (6.9) TeV$^{-4}$
& $\left|\dfrac{\tilde{f}^T_{3}}{\Lambda^4}\right|$
& $\dfrac{2}{5}(3+4\sqrt{3}) \, \pi \, s^{-2}$
& 2.5 (0.15) TeV$^{-4}$
\\[+0.2cm]

$\left|\dfrac{\tilde{f}^M_{4}}{\Lambda^4}\right|$
& $64\sqrt{3} \, \pi \, s^{-2}$
&69 (4.3) TeV$^{-4}$
& $\left|\dfrac{\tilde{f}^T_{4}}{\Lambda^4}\right|$
& $\dfrac{4}{5}(9+2\sqrt{3})\, \pi \, s^{-2}$
& 6.2 (0.39) TeV$^{-4}$
\\[+0.2cm]

$\left|\dfrac{\tilde{f}^M_{5}}{\Lambda^4}\right|$
& $\dfrac{8\sqrt{6}}{3}\, \pi \, s^{-2}$
& 4.1 (0.25) TeV$^{-4}$
& $\left|\dfrac{\tilde{f}^T_{5}}{\Lambda^4}\right|$
& $\dfrac{22}{5} \, \pi \, s^{-2}$
&2.7 (0.17) TeV$^{-4}$
\\[+0.2cm]

$\left|\dfrac{\tilde{f}^M_{6}}{\Lambda^4}\right|$
& $\dfrac{16}{5}(24+\sqrt{6})\, \pi \, s^{-2}$
& 53 (3.3) TeV$^{-4}$
& $\left|\dfrac{\tilde{f}^T_{6}}{\Lambda^4}\right|$
& $\dfrac{26}{5} \, \pi \, s^{-2}$
& 3.2 (0.20) TeV$^{-4}$
\\[+0.2cm]

\hline
\end{tabular}%
}

\vspace{2mm}

\refstepcounter{table}
\label{tab:multi}

\makebox[\textwidth][c]{%
\begin{minipage}{0.95\textwidth}
\small
\justifying
\noindent\textbf{TABLE~\thetable.}
Perturbative unitarity constraints on the C- or P-odd Wilson coefficients $f^M_i$, $\tilde{f}^M_{i}$ and $\tilde{f}^T_{i}$ after
marginalizing over the remaining Wilson coefficients for the second
benchmark.
\end{minipage}%
}

\end{table}

\subsection{Third Benchmark: Positivity-Informed Unitarity Constraints}
\label{sim_ops}

Despite their usefulness in charting the SMEFT landscape,
configurations with non-vanishing CP-odd coefficients but without the necessary CP-even counterparts are generally incompatible with the positivity requirements of a UV completion satisfying the stated assumptions. As hinted at by recent work (cf.\ Refs.~\cite{Remmen:2019cyz, Bresciani:2025ojh}), the lowest-order positivity constraints on Wilson coefficients of the CP-odd operators generally come as inequalities of the form $\tilde{f}^2 \lesssim f_i f_j$ involving specific CP-even coefficients. A common schematic form of these correlations is
\begin{equation}
  L_{\textrm{even}}(\boldsymbol{f})\geq\sqrt{Q_{\textrm{even}}(\boldsymbol{f})+Q_{\textrm{odd}}(\boldsymbol{\tilde{f}})}   
    \;  ,
    \label{fff}
\end{equation}
where $L_{\mathrm{even}}$ is linear and $Q_{\mathrm{even}}$,
$Q_{\mathrm{odd}}$ are quadratic in the corresponding CP-even and
CP-odd Wilson coefficients $\boldsymbol f$ and $\boldsymbol{\tilde f}$. Such relations generally imply that nonzero CP-odd coefficients come accompanied by non-vanishing CP-even ones when they arise from a complete UV theory.\smallskip

The above realization, therefore, prompts us to extend our study to
positivity-informed perturbative unitarity bounds, allowing the simultaneous presence of CP-even and CP-odd operators. As in the
previous benchmark, we can largely simplify the unitarity analysis by reducing the problem to the study of subblocks. However, for
positivity analyses, we retain the separate treatment of the complete mixed and tensorial systems established in Sec.~\ref{sec:positivity}, without including correlations with the purely scalar sector. Within these systems, a full separation of the individual operator sectors is generally not appropriate: the $\Phi^2BW$ ($B^2W^2$) sectors require the inclusion of $\Phi^2B^2$ and $\Phi^2W^2$ ($B^4$, and $W^4$) sectors to generically allow for nonzero Wilson coefficients.\smallskip

For this reason, we consider each of the six electroweak sectors
previously defined, carefully specifying the in-sector and out-of-sector operator contributions within the corresponding mixed or
tensorial system. We further outline the eigenvalues of the coupled-channel scattering matrices used to impose perturbative unitarity, together with the associated forward-limit positivity constraints. To obtain the latter, we employ representative electroweak superposition states detailed in App.~\ref{app:uchoices}. To our knowledge, the resulting necessary positivity relations encompassing all CP-odd aQGC operators have not been presented elsewhere. \smallskip

We note that, for tensorial operators only, all independent Wilson-coefficient directions already enter the $J=0$ partial-wave matrices. In our explicit diagonalization, the additional $J=1$ tensorial eigenvalues do not lead to stronger
constraints than those obtained from the $J=0$ sector. We therefore
use the $J=0$ tensorial eigenvalues to derive the quoted tensorial
unitarity bounds. On another note, the positivity-informed perturbative unitarity bounds of the $B^2W^2$ and $\Phi^2BW$ sectors
require the unitarity eigenvalues of the full tensorial and mixed
sectors, respectively. Since the corresponding expressions are
lengthy, we collect them in App.~\ref{app:eigenvalues}. \smallskip

\subsubsection{$B^4$ operators}
This sector comprises the three tensorial operators ${\cal O}^T_8$, ${\cal O}^T_9$ and ${\cancel{\cal O}^T_1}$. For $J=0$, only the $Q=0$ matrix is non-vanishing and has the two independent eigenvalues
\begin{equation}
    \frac{s^2}{16\pi\Lambda^4}
    \left(
    4 f^T_{8}+3 f^T_{9}
    \pm \frac{5}{3}
    \sqrt{\left(4 f^T_{8}+f^T_{9}\vphantom{\tilde f}\right)^2+\left(4\tilde f^T_{1}\right)^2}
    \right) \; .
\end{equation}
The positivity bound for this sector showcases the similar structure
\begin{equation}
    4 f^T_8+3f^T_9-\sqrt{ \left(4 f^T_{8} + f^T_{9} \vphantom{\tilde f}\right)^2+ \left(4\tilde{f}^T_{1}\right)^2}\geq 0 \; .
\end{equation}
This relation exactly reproduces the bounds reported in Eq.~(42) in Ref.~\cite{Remmen:2019cyz}.

\subsubsection{$W^4$ operators}

This sector comprises the six tensorial operators ${\cal O}^T_{0}$,
${\cal O}^{T}_{1}$, ${\cal O}^T_2$, ${\cal O}^T_{3}$, ${\cancel{\cal O}^{T}_{5}}$, and ${\cancel{\cal O}^T_6}$. For $J=0$,
the $Q=0$ matrix provides a complete set of independent eigenvalues to fully establish perturbative unitarity bounds, while the $Q=1$ and $Q=2$ matrices reproduce only the first two eigenvalues:
\begin{align}
    &\dfrac{s^2}{32\pi \Lambda^4}\left(2 f^T_{1}+f^T_{2}+f^T_{3} \pm \dfrac{1}{3}\sqrt{\left(4f^T_{0} + 8 f^T_{1} + f^T_{2} + 3 f^T_{3}\vphantom{\tilde f}\right)^2 + 256  \left(\tilde{f}^T_{5} + 2 \tilde{f}^T_{6}\right)^2}\right) \; ,\label{w4eigen}\\[2mm]
    &\dfrac{s^2}{64\pi \Lambda^4}\left(12 f^T_{0} + 4f^T_{1} + 5\left(f^T_{2} + f^T_{3}\vphantom{\tilde f}\right)\pm\dfrac{1}{3}\sqrt{\left[ 11\left(4f^T_{0}+f^T_{2}\vphantom{\tilde f}\right) + 28 f^T_{1}+ 3f^T_{3} \right]^2 + 256 \left(11\tilde{f}^T_{5} + 7 \tilde{f}^T_{6}\right)^2}\right) \; .
\end{align}
The relevant quadratic positivity bounds are:
\begin{align}
    &2 f^T_1+f^T_2+f^T_3-\sqrt{\left( 2f^T_1+f^T_3\vphantom{\tilde f}\right)^2 + \left( 8\tilde{f}^T_{6} \right)^2}\geq 0 \; ,\\[2mm]
    &2 f^T_0+f^T_1+f^T_2+f^T_3-\sqrt{\left( 2f^T_0+f^T_1+\dfrac{1}{2}f^T_2\vphantom{\tilde f}\right)^2 + \left( 8\tilde{f}^T_{5} +4\tilde{f}^T_6 \right)^2}\geq 0 \; .
    \label{remmenw42}
\end{align}
The bounds reported in Eq.~(40) of Ref.~\cite{Remmen:2019cyz} are
recovered from the factorized weak-isospin--helicity superpositions. Including the non-factorizable direction defined in
App.~\ref{app:uchoices} yields the stronger constraint in Eq.~\eqref{remmenw42}, further restricting the allowed Wilson-coefficient parameter space beyond the bounds of Ref.~\cite{Remmen:2019cyz}. \smallskip

\subsubsection{$B^2 W^2$ operators}

This sector comprises the seven tensorial operators ${\cal O}^T_4$,
${\cal O}^T_5$, ${\cal O}^T_6$, ${\cal O}^T_7$,
${\cancel{\cal O}^T_2}$, ${\cancel{\cal O}^T_3}$, and
${\cancel{\cal O}^T_4}$. As in the $W^4$ sector, the ($J=0$, $Q=0$)
channel provides the complete set of independent eigenvalues. The
addition of the ($J=0$, $Q=1$) matrix reproduces the first two
eigenvalues, while the $Q=2$ matrix vanishes. Altogether, we obtain
the following unitarity eigenvalues
\begin{align}
    &\dfrac{s^2}{32\pi \Lambda^4}\left(f^T_{4}+2f^T_{6}+f^T_{7} \pm \dfrac{1}{3}\sqrt{\left( 3f^T_{4} + 4 f^T_{5} + 8f^T_{6} + f^T_{7}\vphantom{\tilde f}\right)^2 + 64 \left(\tilde{f}^T_{2} +\tilde{f}^T_{3}+2\tilde{f}^T_{4}\right)^2}\right) \; ,\label{BWeigen}\\[2mm]
    &\dfrac{\pm\, s^2}{32\sqrt{2}\pi \Lambda^4}\sqrt{A_1+A_2\pm\sqrt{A_3 A_4}} \; ,
\end{align}
with
\begin{align}
    A_1&= \dfrac{5}{3} \left(f^T_{4}\vphantom{\tilde f}\right)^2 + 8 \left[\dfrac{1}{3}\left( 3f^T_{5}+ f^T_{6}\vphantom{\tilde f}\right)^2 +3\left(f^T_5\vphantom{\tilde f}\right)^2\right]+ 4 \left(6 f^T_{5}+f^T_{6}\vphantom{\tilde f}\right) f^T_{7}+ 3 \left(f^T_{7}\vphantom{\tilde f}\right)^2+ 2f^T_{4}\left[4 f^T_{5}-\dfrac{2}{3} f^T_{6}+f^T_{7}\right],\\[2mm]
    A_2&=\dfrac{16}{3} \left[\left(6\tilde{f}^T_{2}+\tilde{f}^T_{4}\right)^2+\left(6\tilde{f}^T_{3}+\tilde{f}^T_{4}\right)^2\right] \, ,\\[2mm]
    A_3&=64 \left[3\left(\tilde{f}^T_{2}+\tilde{f}^T_{3}\right)+\tilde{f}^T_{4}\right]^2+\left[f^T_{4}-4\left(3f^T_{5}+f^T_{6}\vphantom{\tilde f}\right)-3{f^T_{7}}\vphantom{\tilde f}\right]^2 \, ,\\[2mm]
    A_4&=64 \left(\tilde{f}^T_{2}-\tilde{f}^T_{3}\right)^2+\left(f^T_{4}+4f^T_{5}+f^T_{7}\vphantom{\tilde f}\right)^2.
\end{align}

The positivity bounds for the $B^2W^2$ operators must be derived in
the combined tensorial system that includes the $B^4$ and $W^4$
sectors. Therefore, in addition to the positivity conditions obtained from the $B^4$ and $W^4$ sectors, the $B^2W^2$ sector imposes
\begin{align}
    &f^T_4+2f^T_6+f^T_7-\sqrt{\left( f^T_4+2f^T_6\vphantom{\tilde f}\right)^2 +\left( 4\tilde{f}^T_{4}\right)^2}\geq 0 \; ,\label{remmenb2w2}\\[2mm]
    &2f^T_0+f^T_1+f^T_2+f^T_3+f^T_4+2f^T_6+f^T_7+2\left(4f^T_8+3f^T_9\vphantom{\tilde f}\right)-\sqrt{A_5}\geq 0 \; ,\\[2mm]
    &4\left(f^T_0+f^T_1\vphantom{\tilde f}\right)+3\left(f^T_2+f^T_3\vphantom{\tilde f}\right)+2\left(f^T_4+2f^T_6+f^T_7\vphantom{\tilde f}\right)+4\left(4f^T_8+3f^T_9\vphantom{\tilde f}\right)-\sqrt{A_6}\geq 0 \; ,\\[2mm]
    &4\left(f_0^T+2f_1^T\vphantom{\tilde f}\right)+5\left(f_2^T+f_3^T\vphantom{\tilde f}\right)+4\left(f_4^T+2f_6^T+f_7^T\vphantom{\tilde f}\right)+4\left(4f_8^T+3f_9^T\vphantom{\tilde f}\right)-\sqrt{A_7}\geq 0 \; ,
\end{align}
where
\begin{align}
    A_5&=\left[2f^T_0+f^T_1+f^T_2+f^T_3-2\left(4f^T_8+3f^T_9\vphantom{\tilde f}\right)\right]^2+\left(4f^T_5+2f^T_6+f^T_7\vphantom{\tilde f}\right)^2+16\left[2\left(\tilde{f}^T_2+\tilde{f}^T_3\right)+\tilde{f}^T_4\right]^2 \; ,\\[2mm]
    A_6&=\Big[4\left(f^T_0+f^T_1\vphantom{\tilde f}\right)+3\left(f^T_2+f^T_3\vphantom{\tilde f}\right)-4\left(4f^T_8+3f^T_9\vphantom{\tilde f}\right)\Big]^2+4\Big[f^T_4+2f^T_6+f^T_7+\left|f^T_4+4f^T_5+f^T_7\right|\Big]^2+\nonumber\\[1mm]
    &\hspace{5mm}+256\left(\tilde{f}^T_2-\tilde{f}^T_3\right)^2 \; ,\\[2mm]
    A_7&=\Big[4\left(f_0^T+2f_1^T\vphantom{\tilde f}\right)+5\left(f_2^T+f_3^T\vphantom{\tilde f}\right)+2\left(f_4^T+2f_6^T+f_7^T\vphantom{\tilde f}\right)-4\left(4f_8^T+3f_9^T\vphantom{\tilde f}\right)\Big]^2+\nonumber\\[1mm]
    &\hspace{5mm}+\Big[f_4^T+2f_6^T+f_7^T+2\left|f_4^T+4f_5^T+f_7^T\right|\Big]^2+\Big[\sqrt{3}\left(f_4^T+2f_6^T+f_7^T\vphantom{\tilde f}\right)+16\big|\tilde{f}_2^{T}-\tilde{f}_3^{T}\big|\Big]^2 \; .
\end{align}

The bound in Eq.~(\ref{remmenb2w2}) exactly reproduces the first three lines of Eq.~(46) in Ref.~\cite{Remmen:2019cyz}. As a reminder, the complete tensorial-sector eigenvalues used to derive these positivity-improved unitarity bounds are collected in App.~\ref{app:eigenvalues}.

\subsubsection{$\Phi^2B^2$ operators}

The $\Phi^2B^2$ sector contains the operators ${\cal O}_2^M$, ${\cal O}_3^M$ and ${\cancel{\cal O}}_1^M$. For $J=0$, only the $Q=0$ matrix is non-vanishing. For $J=1$, the $Q=0$ and $Q=1$ matrices yield the same independent eigenvalues, while the $Q=2$ matrix vanishes. Collecting the nonzero eigenvalues from both partial waves, we obtain
\begin{align}
    &\textbf{\textit{J}\hspace{0.3mm}=\hspace{0.8mm}0}: \qquad\qquad\dfrac{\pm\,s^2}{32\sqrt{2}\pi \Lambda^4}\sqrt{\left( 4 f^M_{2} - f^M_{3} \vphantom{\tilde f}\right)^2+ \left( 4\tilde{f}^M_{1} \right)^2} \; .\\[2mm]
    &\textbf{\textit{J}\hspace{0.3mm}=\hspace{0.8mm}1}:\qquad\qquad
    \dfrac{s^2}{256\pi \Lambda^4}\left(\dfrac{8}{3}f_3^M\pm\sqrt{\left( 4 f^M_{2} - f^M_{3} \vphantom{\tilde f}\right)^2+ \left( 4\tilde{f}^M_{1} \right)^2}\right) \; .
\end{align}
The corresponding positivity bound reads
\begin{equation}
    f^M_3-\sqrt{ \left(4 f^M_{2}-f^M_{3} \vphantom{\tilde f}\right)^2+ \left( 4\tilde{f}^M_{1} \right)^2}\geq 0 \, .
  \end{equation}
This relation coincides with Eq.~(5.21) of Ref.~\cite{Bresciani:2025ojh} after the substitution $X=B$.

\subsubsection{$\Phi^2W^2$ operators}

In this sector there are six mixed operators ${\cal O}^M_0$, ${\cal O}^M_1$, ${\cal O}^M_7$, ${\cal O}^M_9$, ${\cancel{\cal O}^M_5}$, and ${\cancel{\cal O}^M_6}$. For $J=0$, only the $Q=0$ matrix is non-vanishing and provides non-vanishing eigenvalues. For $J=1$, the $Q=0$ and $Q=1$ scattering matrices provide the full set of independent eigenvalues, while the $Q=2$ matrix only gives rise to the second set displayed below. The resulting eigenvalues are
\begin{align}
    &\textbf{\textit{J}\hspace{0.3mm}=\hspace{0.8mm}0}: \qquad\qquad\dfrac{\pm\,\sqrt{6}s^2}{256\pi \Lambda^4}\sqrt{\left( 8f^M_0-2f^M_1+f^M_7 \vphantom{\tilde f}\right)^2+ \left( 16\tilde{f}^M_{5} \right)^2} \; .\\[2mm]
    &\textbf{\textit{J}\hspace{0.3mm}=\hspace{0.8mm}1}:\qquad\qquad
    \dfrac{\pm\, s^2}{384\pi \Lambda^4}\sqrt{\left( f^M_7 \vphantom{\tilde f}\right)^2+ \left( 4\tilde{f}^M_{6} \right)^2} \; ,\\[2mm]
    &\hspace{26mm}\dfrac{s^2}{384\pi \Lambda^4}\left(2f^M_1-f^M_7\pm\dfrac{1}{4}\sqrt{\left(12 f^M_0-3f^M_1+4f^M_7\vphantom{\tilde f}\right)^2+4 \left[\left( 12\tilde{f}^M_{5}-5\tilde{f}^M_6 \vphantom{\tilde f}\right)^2+\left( 8 f^M_{9}\vphantom{\tilde f}\right)^2\right]}\right) \; ,\label{j1hhww}\\[2mm]
    &\hspace{26mm}\dfrac{s^2}{384\pi \Lambda^4}\left(2f^M_1-f^M_7\pm\dfrac{1}{8}\sqrt{\left(24 f^M_0-6f^M_1-7f^M_7\vphantom{\tilde f}\right)^2+64 \left[\left( 6\tilde{f}^M_{5}+5\tilde{f}^M_6 \right)^2+\left( 8 f^M_{9}\vphantom{\tilde f}\right)^2\right]}\right) \; .
\end{align}
The relevant linear and quadratic positivity bounds are:
\begin{align}
  &2f^M_1-f^M_7-\frac{2}{3}\left(2+\sqrt{3}+\sqrt{31-4\sqrt{3}}\right)\left|f^M_9\right|\geq 0,\\[2mm]
  &2f^M_1-f^M_7-\sqrt{\left( f^M_7\vphantom{\tilde f}\right)^2+\left(4f^M_9\vphantom{\tilde f}\right)^2 +\left(4 \tilde{f}^M_{6}\right)^2}\geq 0 \; ,\\[2mm]
  &2f^M_1-f^M_7-\sqrt{\left( 8f^M_0-2f^M_1+f^M_7\vphantom{\tilde f}\right)^2 +\left(4 f^M_9\vphantom{\tilde f}\right)^2+\left( 16\tilde{f}^M_{5}\right)^2}\geq 0,\label{brescianih2w2}\\[2mm]
  &2f^M_1-f^M_7-4\left|f^M_9\right|-\sqrt{\left( 4f^M_0-f^M_1\vphantom{\tilde f}\right)^2 +4\left( 4\tilde{f}^M_{5}+\tilde{f}^M_6\right)^2}\geq 0 \; .
\end{align}
In particular, we have checked that Eq.~(\ref{brescianih2w2})
corresponds to Eq.~(5.21) of Ref.~\cite{Bresciani:2025ojh} for $X=W$
and in the limit $f^M_9=0$. For non-vanishing values of $f^M_9$, our
expression provides a more stringent constraint. \smallskip

\subsubsection{$\Phi^2BW$ operators}

This sector comprises the six mixed operators ${\cal O}^M_4$,
${\cal O}^M_5$, ${\cal O}^M_8$, ${\cancel{\cal O}^M_2}$,
${\cancel{\cal O}^M_3}$, and ${\cancel{\cal O}^M_4}$. For each fixed
value of $J=0,1$, both the $Q=0$ and $Q=1$ matrices provide the same
independent eigenvalues, while the $Q=2$ matrix vanishes. The
unitarity eigenvalues are
\begin{align}
    &\textbf{\textit{J}\hspace{0.3mm}=\hspace{0.8mm}0}: \qquad\qquad\dfrac{\pm\,s^2}{64\pi \Lambda^4}\sqrt{\left( 2f^M_4+f^M_5 \vphantom{\tilde f}\right)^2+ \left( 2\tilde{f}^M_{2}+\tilde{f}^M_4 \right)^2} \; .\label{j0hhbw}\\[2mm]
    &\textbf{\textit{J}\hspace{0.3mm}=\hspace{0.8mm}1}:\qquad\qquad
    \dfrac{\pm\,s^2}{192\pi \Lambda^4}\sqrt{\left( f^M_8 \vphantom{\tilde f}\right)^2+ \left( \tilde{f}^M_{3} \right)^2} \; ,\\[2mm]
    &\hspace{26mm}\dfrac{\pm\,s^2}{256\sqrt{2}\pi \Lambda^4}\sqrt{B_1 \pm \sqrt{B_2}} \; ,
\end{align}
with
\begin{align}
    B_1&=\dfrac{1}{6}\left[\left( 8 f^M_5\vphantom{\tilde f}\right)^2+\left(5f^M_8\vphantom{\tilde f}\right)^2+\left(6 f^M_4 + 3 f^M_5\vphantom{\tilde f}\right)^2+\left(8 \tilde{f}^M_4\right)^2 +\left(5\tilde{f}^M_3\right)^2+\left(6 \tilde{f}^M_2 + 3\tilde{f}^M_4\right)^2\right] \; ,\\[2mm]
    B_2&=25\left[\left(2f_4^M+f_5^M\vphantom{\tilde
         f}\right)f_8^M-\tilde f_3^M\left(2\tilde f_2^M+\tilde
         f_4^M\right)\right]^2+64\left[\left(f_5^M\vphantom{\tilde
         f}\right)^2+\left(\tilde
         f_4^M\right)^2\right]\left[\left(2f_4^M+f_5^M\vphantom{\tilde
         f}\right)^2+\left(2\tilde f_2^M+\tilde
         f_4^M\right)^2\right]\; .
\end{align}
As in the case of the $B^2W^2$ sector, the positivity bounds for the operators of the $\Phi^2BW$ sector must be derived in the combined mixed system that includes the $\Phi^2B^2$ and $\Phi^2W^2$ sectors. Therefore, in addition to the positivity conditions obtained from the $\Phi^2B^2$ and $\Phi^2W^2$ sectors, the $\Phi^2BW$ sector requires that
\begin{align}
    &2f^M_3 \left(2f^M_1-f^M_7\vphantom{\tilde f}\right)\geq2\left(f^M_8\vphantom{\tilde f}\right)^2+2\left(\tilde{f}^M_3\right)^2+\left(2f^M_4+f^M_5\vphantom{\tilde f}\right)^2+\left(f^M_5\vphantom{\tilde f}\right)^2+\left(2\tilde{f}^M_2+\tilde{f}^M_4\right)^2+\nonumber\\[1mm]
    &\hspace{33mm}+\left(\tilde{f}^M_4\right)^2+\left|\left(2 f^M_4+f^M_5\vphantom{\tilde f}\right)^2-\left(f^M_5\vphantom{\tilde f}\right)^2+\left(2\tilde{f}^M_2+\tilde{f}^M_4\right)^2-\left(\tilde{f}^M_4\right)^2\right| \; ,\\[2mm]   
    &2f^M_3\left(2f^M_1-f^M_7-4\left|f^M_9\right|\vphantom{\tilde f}\right)\geq \left(2 f^M_4+f^M_5\vphantom{\tilde f}\right)^2+\left(f^M_8\vphantom{\tilde f}\right)^2+\left(2\tilde{f}^M_2+\tilde{f}^M_4\right)^2+\left(\tilde{f}^M_3\right)^2+\nonumber\\[1mm]
    &\hspace{47mm}+2\left|f^M_8\left(2 f^M_4+f^M_5\vphantom{\tilde f}\right)-\tilde{f}^M_3\left(2\tilde{f}^M_2+\tilde{f}^M_4\right)\right| \; ,\\[2mm]  
    &\frac{1}{2}\left(2f^M_1+4f^M_3-f^M_7\vphantom{\tilde f}\right)-\left|f^M_9\right|-\sqrt{\left(f^M_9\vphantom{\tilde f}\right)^2+2\left(\left(f^M_5\vphantom{\tilde f}\right)^2+\left(\tilde{f}^M_4\vphantom{\tilde f}\right)^2+\left|\left(f^M_5\vphantom{\tilde f}\right)^2-\left(\tilde{f}^M_4\vphantom{\tilde f}\right)^2\right|\right)}\geq 0 \; ,\\[2mm]    
    &\frac{3}{4}\left(2f^M_1+4f^M_3-f^M_7\vphantom{\tilde f}\right)-\left|f^M_9\right|-\sqrt{B_3}\geq 0 \; ,\\[2mm]      
    &\frac{3}{2}\left(2f^M_1+4f^M_3-f^M_7\vphantom{\tilde f}\right)-\left(2+\sqrt{3}\right)\left|f^M_9\right|-\sqrt{B_4}\geq 0 \; ,\\[2mm]
    &f^M_3\left(2f^M_1-f^M_7-|f^M_9|\vphantom{\tilde f}\right)-\left(f^M_5\vphantom{\tilde f}\right)^2-\left(\tilde{f}^M_4\vphantom{\tilde f}\right)^2\geq 0 \; ,
\end{align}
with
\begin{align}
    B_3&=\left(f^M_9\vphantom{\tilde f}\right)^2\!+\frac{9}{2}\left[\left(\tilde{f}^M_4\right)^2\!+\left(f^M_5\vphantom{\tilde f}\right)^2\!+\left|\left(\tilde{f}^M_4\right)^2-\left(f^M_5\vphantom{\tilde f}\right)^2\right|\right]\!+\sqrt{6}\left|f^M_9\right|\sqrt{\left(\tilde{f}^M_4\right)^2\!+\left(f^M_5\vphantom{\tilde f}\right)^2\!+\left|\left(\tilde{f}^M_4\right)^2-\left(f^M_5\vphantom{\tilde f}\right)^2\right|} \; ,\\[2mm]
    B_4&=18\left[\left(\tilde{f}^M_4\right)^2+\left(f^M_5\vphantom{\tilde f}\right)^2\right]+\left(31-4\sqrt{3}\right)\left(f^M_9\vphantom{\tilde f}\right)^2+\nonumber\\[1mm]
    &\hspace{5mm}+2\left(3+2\sqrt{3}\right)\sqrt{2\left(f^M_9\vphantom{\tilde f}\right)^2\left[\left(f^M_5\vphantom{\tilde f}\right)^2\!+\left(\tilde{f}^M_4\right)^2\right]\!+\!\left|\left(f^M_9\vphantom{\tilde f}\right)^2\left[\left(f^M_5\vphantom{\tilde f}\right)^2\!\!-\left(\tilde{f}^M_4\right)^2\right]\!-2\sqrt{3}~f^M_5 \left(f^M_9\left|f^M_9\right|\vphantom{\tilde f}\right)\tilde{f}^M_4\right|} \; .
\end{align}
As a reminder, the complete mixed-sector eigenvalues used to derive these positivity-improved unitarity bounds are collected in App.~\ref{app:eigenvalues}.

\begin{table}[h]

\setbox0=\hbox{%
\begin{tabular}{|c|c|c||c|c|c|}
\hline
 & \multicolumn{2}{c||}{\rule{0pt}{2.5ex}Bounds on $f^M_{i}$} 
 & 
 & \multicolumn{2}{c|}{\rule{0pt}{2.5ex}Bounds on $f^T_{i}$} 
\\ \cline{2-3}\cline{5-6}

\shortstack{Wilson\\coefficient}
& Analytical
& For $\sqrt{s}<1.5\, (3)$ TeV
& \shortstack{Wilson\\coefficient}
& Analytical
& For $\sqrt{s}<1.5\, (3)$ TeV
\\
\hline
&&&&&\\[-0.2cm]

$\left|\dfrac{f^M_{0}}{\Lambda^4}\right|$
& $\dfrac{2}{3}\left(72+5\sqrt{6}\right)\, \pi \, s^{-2}$
& 35 (2.2) TeV$^{-4}$
& $\color{blue}\left|\dfrac{f^T_{0}}{\Lambda^4}\right|$
& $\color{blue}12\, \pi \, s^{-2}$
& \color{blue}7.4 (0.47) TeV$^{-4}$
\\[+0.2cm]

$\left|\dfrac{f^M_{1}}{\Lambda^4}\right|$
& $\dfrac{8}{5}\left(120+\sqrt{6}\right)\, \pi \, s^{-2}$
& 122 (7.6) TeV$^{-4}$
& $\color{blue}\left|\dfrac{f^T_{1}}{\Lambda^4}\right|$
& $\color{blue}8\, \pi \, s^{-2}$
& \color{blue}5.0 (0.31) TeV$^{-4}$
\\[+0.2cm]

$\left|\dfrac{f^M_{2}}{\Lambda^4}\right|$
& $\left(24+5\sqrt{2}\right)\, \pi \, s^{-2}$
& 19 (1.2) TeV$^{-4}$
& $\color{blue}\left|\dfrac{f^T_{2}}{\Lambda^4}\right|$
& $\color{blue}\dfrac{704}{19}\, \pi \, s^{-2}$
& \color{blue}23 (1.4) TeV$^{-4}$
\\[+0.2cm]

$\left|\dfrac{f^M_{3}}{\Lambda^4}\right|$
& $96\, \pi \, s^{-2}$
& 60 (3.7) TeV$^{-4}$
& $\color{blue}\left|\dfrac{f^T_{3}}{\Lambda^4}\right|$
& $\color{blue}\color{blue}\dfrac{352}{19}\, \pi \, s^{-2}$
& \color{blue}11 (0.72) TeV$^{-4}$
\\[+0.2cm]

$\color{blue}\left|\dfrac{f^M_{4}}{\Lambda^4}\right|$
& $\color{blue}\beta_{24}\, \pi \, s^{-2}$
& \color{blue}42 (2.6) TeV$^{-4}$
& $\color{blue}\left|\dfrac{f^T_{4}}{\Lambda^4}\right|$
& $\color{blue}\alpha_4\, \pi \, s^{-2}$
& \color{blue}16 (0.99) TeV$^{-4}$
\\[+0.2cm]

$\color{blue}\left|\dfrac{f^M_{5}}{\Lambda^4}\right|$
& $\color{blue}96\left(\sqrt{3}-1\right)\, \pi \, s^{-2}$
& \color{blue}44 (2.7) TeV$^{-4}$
& $\color{blue}\left|\dfrac{f^T_{5}}{\Lambda^4}\right|$
& $\color{blue}\dfrac{4}{15}\left(27+\sqrt{3\,\alpha_5}\right) \, \pi \, s^{-2}$
& \color{blue}7.6 (0.48) TeV$^{-4}$
\\[+0.2cm]

$//$
& $//$
& $//$
& $\color{blue}\left|\dfrac{f^T_{6}}{\Lambda^4}\right|$
& $\color{blue}\alpha_6 \, \pi \, s^{-2}$
& \color{blue}8.6 (0.54) TeV$^{-4}$
\\[+0.2cm]

$\color{blue}\left|\dfrac{f^M_{7}}{\Lambda^4}\right|$
& $\color{blue}\dfrac{64}{9}\left(24+\sqrt{6}\right)\, \pi \, s^{-2}$
& \color{blue}117 (7.3) TeV$^{-4}$
& $\left|\dfrac{f^T_{7}}{\Lambda^4}\right|$
& $\dfrac{32}{5}\left(5+\sqrt{3}\right)\,\pi\, s^{-2}$
& 27 (1.7) TeV$^{-4}$
\\[+0.2cm]

$\color{blue}\left|\dfrac{f^M_{8}}{\Lambda^4}\right|$
& $\color{blue}\dfrac{1536}{11}\left(5\sqrt{3}-8\right)\, \pi \, s^{-2}$
& \color{blue}57 (3.6) TeV$^{-4}$
& $\color{blue}\left|\dfrac{f^T_{8}}{\Lambda^4}\right|$
& $\color{blue}3\, \pi \, s^{-2}$
& \color{blue}1.9 (0.12) TeV$^{-4}$
\\[+0.2cm]

$\color{blue}\left|\dfrac{f^M_{9}}{\Lambda^4}\right|$
& $\color{blue}\dfrac{576}{14+\sqrt{3}+\sqrt{31-4\sqrt{3}}}\, \pi \, s^{-2}$
& \color{blue}17 (1.1) TeV$^{-4}$
& $\left|\dfrac{f^T_{9}}{\Lambda^4}\right|$
& $8\, \pi \, s^{-2}$
& 5.0 (0.31) TeV$^{-4}$
\\[+0.35cm]

\hline
\end{tabular}%
}

\dimen0=\wd0

\makebox[\textwidth][c]{\box0}

\vspace{2mm}

\refstepcounter{table}
\label{tab:multi2}

\makebox[\textwidth][c]{%
  \begin{minipage}{0.95\textwidth} \small \justifying
    \noindent\textbf{TABLE~\thetable.} Positivity-improved
    perturbative unitarity constraints on the Wilson coefficients of
    the CP-even operators ${\cal O}^{M}_i$ and ${\cal O}^{T}_i$ for
    the third benchmark, {\em i.e.} with multiple operators simultaneously active. Bounds tightened by positivity are highlighted in blue. The polynomials that yield the numerical constants $\alpha_4\approx 25.433$, $\alpha_5\approx 121.837$, $\alpha_6\approx 13.937$ and $\beta_{24}\approx 66.971$ are provided in App.~\ref{app:polynomials}.
  \end{minipage}%
} \end{table}

\begin{table}

\setbox0=\hbox{%
\begin{tabular}{|c|c|c||c|c|c|}
\hline
 & \multicolumn{2}{c||}{\rule{0pt}{2.5ex}Bounds on $\tilde{f}^M_{i}$} 
 & 
 & \multicolumn{2}{c|}{\rule{0pt}{2.5ex}Bounds on $\tilde{f}^T_{i}$} 
\\ \cline{2-3}\cline{5-6}

\shortstack{Wilson\\coefficient}
& Analytical
& For $\sqrt{s}<1.5\, (3)$ TeV
& \shortstack{Wilson\\coefficient}
& Analytical
& For $\sqrt{s}<1.5\, (3)$ TeV
\\
\hline
&&&&&\\[-0.2cm]

$\left|\dfrac{\tilde{f}^M_{1}}{\Lambda^4}\right|$
& $8\sqrt{2}\, \pi \, s^{-2}$
& 7.0 (0.44) TeV$^{-4}$
& $\color{blue}\left|\dfrac{\tilde{f}^T_{1}}{\Lambda^4}\right|$
& $\color{blue}\dfrac{3}{2}\, \pi \, s^{-2}$
& \color{blue}0.93 (0.058) TeV$^{-4}$
\\[+0.2cm]

$\color{blue}\left|\dfrac{\tilde{f}^M_{2}}{\Lambda^4}\right|$
& $\color{blue}\beta_{24}\, \pi \, s^{-2}$
& \color{blue}42 (2.6) TeV$^{-4}$
& $\color{blue}\left|\dfrac{\tilde{f}^T_{2}}{\Lambda^4}\right|$
& $\color{blue}\widetilde{\alpha}_{23}\, \pi \, s^{-2}$
& \color{blue}2.0 (0.12) TeV$^{-4}$
\\[+0.2cm]

$\color{blue}\left|\dfrac{\tilde{f}^M_{3}}{\Lambda^4}\right|$
& $\color{blue}\dfrac{1536}{11}\left(5\sqrt{3}-8\right)\, \pi \, s^{-2}$
& \color{blue}57 (3.6) TeV$^{-4}$
& $\color{blue}\left|\dfrac{\tilde{f}^T_{3}}{\Lambda^4}\right|$
& $\color{blue}\widetilde{\alpha}_{23}\, \pi \, s^{-2}$
& \color{blue}2.0 (0.12) TeV$^{-4}$
\\[+0.2cm]

$\color{blue}\left|\dfrac{\tilde{f}^M_{4}}{\Lambda^4}\right|$
& $\color{blue}96\left(\sqrt{3}-1\right)\, \pi \, s^{-2}$
& \color{blue}44 (2.7) TeV$^{-4}$
& $\color{blue}\left|\dfrac{\tilde{f}^T_{4}}{\Lambda^4}\right|$
& $\color{blue}\dfrac{1}{403}\left(144+\widetilde{\alpha}_{4}\right)\, \pi \, s^{-2}$
& \color{blue}3.2 (0.20) TeV$^{-4}$
\\[+0.2cm]

$\left|\dfrac{\tilde{f}^M_{5}}{\Lambda^4}\right|$
& $\dfrac{8\sqrt{6}}{3}\, \pi \, s^{-2}$
& 4.1 (0.25) TeV$^{-4}$
& $\color{blue}\left|\dfrac{\tilde{f}^T_{5}}{\Lambda^4}\right|$
& $\color{blue}\dfrac{40}{19}\, \pi \, s^{-2}$
& \color{blue}1.3 (0.082) TeV$^{-4}$
\\[+0.2cm]

$\color{blue}\left|\dfrac{\tilde{f}^M_{6}}{\Lambda^4}\right|$
& $\color{blue}\dfrac{16}{9}(24+\sqrt{6})\, \pi \, s^{-2}$
& \color{blue}29 (1.8) TeV$^{-4}$
& $\color{blue}\left|\dfrac{\tilde{f}^T_{6}}{\Lambda^4}\right|$
& $\color{blue}\dfrac{44}{19}\, \pi \, s^{-2}$
& \color{blue}1.4 (0.090) TeV$^{-4}$
\\

\hline
\end{tabular}%
}

\dimen0=\wd0

\makebox[\textwidth][c]{\box0}

\vspace{2mm}

\refstepcounter{table}
\label{tab:multi3}

\makebox[\textwidth][c]{%
\begin{minipage}{0.95\textwidth}
\small
\justifying
\noindent\textbf{TABLE~\thetable.}
Perturbative unitarity and positivity constraints on the Wilson
coefficients of the CP-odd operators $\cancel{\cal O}^{M}_i$ and
$\cancel{\cal O}^{T}_i$  for the third benchmark. Bounds tightened by positivity are highlighted in blue. The polynomials that yield the numerical constants $\widetilde{\alpha}_{23} \approx 3.179$, $\widetilde{\alpha}_4
\approx1963.873$  and $\beta_{24}\approx 66.971$ are provided
in App.~\ref{app:polynomials}.
\end{minipage}%
}

\end{table}

\subsubsection{Explicit constraints}

Let us combine the above partial-wave unitarity and positivity bounds. When obtaining the marginalized unitarity limits, the inclusion of positivity constraints can reduce the available parameter space. Consequently, the number of possible combinations of couplings that mitigate the rapid growth of the amplitudes is reduced, leading to more stringent limits. We present in Table~\ref{tab:multi2} the marginalized unitarity and positivity constraints on CP-even aQGCs, where we can see that the unitarity limits on most of the couplings are tightened; see the bounds in blue. The corresponding results for the CP-odd operators are
presented in Table~\ref{tab:multi3}. As can be seen from this table, all bounds are more stringent except for the Wilson coefficients $\tilde{f}^M_1$ and $\tilde{f}^M_5$.\smallskip

The analytical bounds for the $B^2W^2$ and $\Phi^2BW$ sectors in Tables~\ref{tab:multi2}--\ref{tab:multi3} involve numerical constants $\alpha_i$, $\widetilde{\alpha}_j$ and $\beta_{24}$, for $i=4,5,6$ and $j=4,23$. These constants are defined as the largest positive real roots of high-degree polynomials, except for $\beta_{24}$, which is defined as the smallest positive real root of its corresponding polynomial. The polynomials are listed in App.~\ref{app:polynomials}.\smallskip

\begin{figure*}[t]

    \makebox[\textwidth][c]{%
    \includegraphics[width=0.85\textwidth]{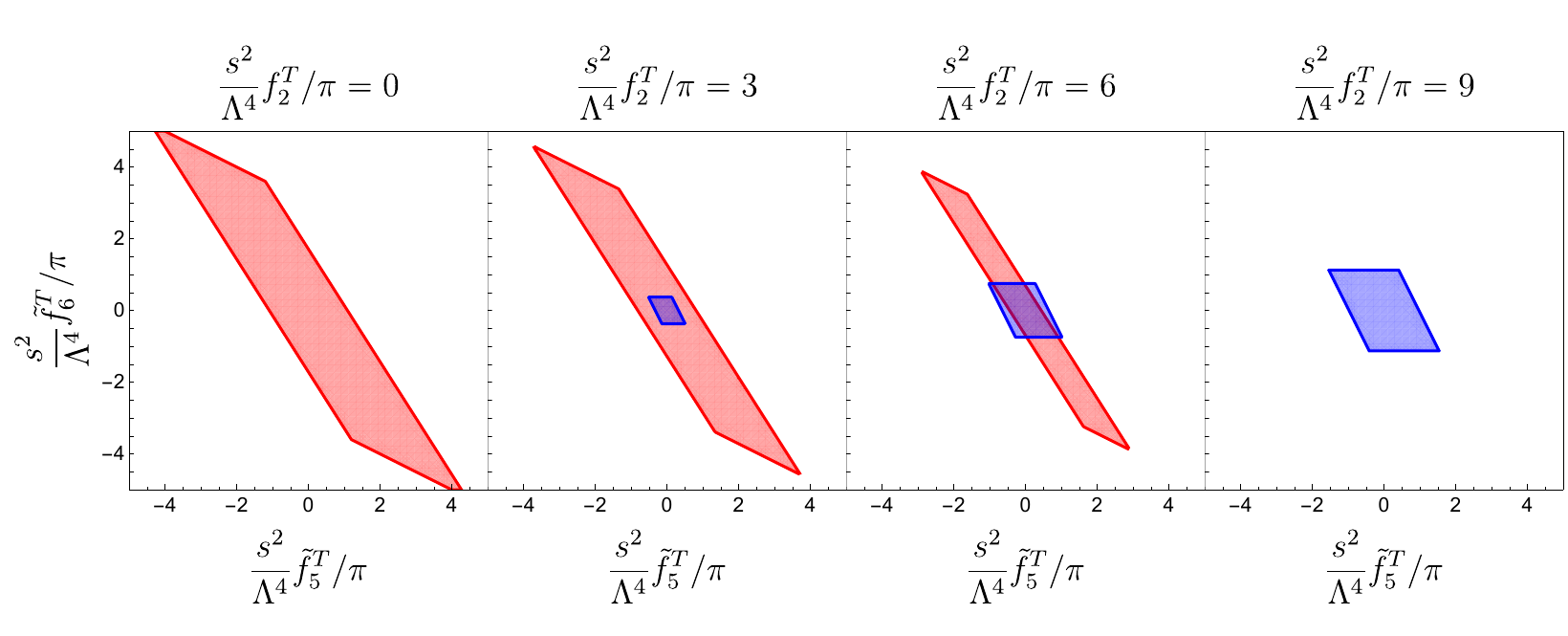}
    }

    \vspace{3mm}

    \makebox[\textwidth][c]{%
    \includegraphics[width=0.85\textwidth]{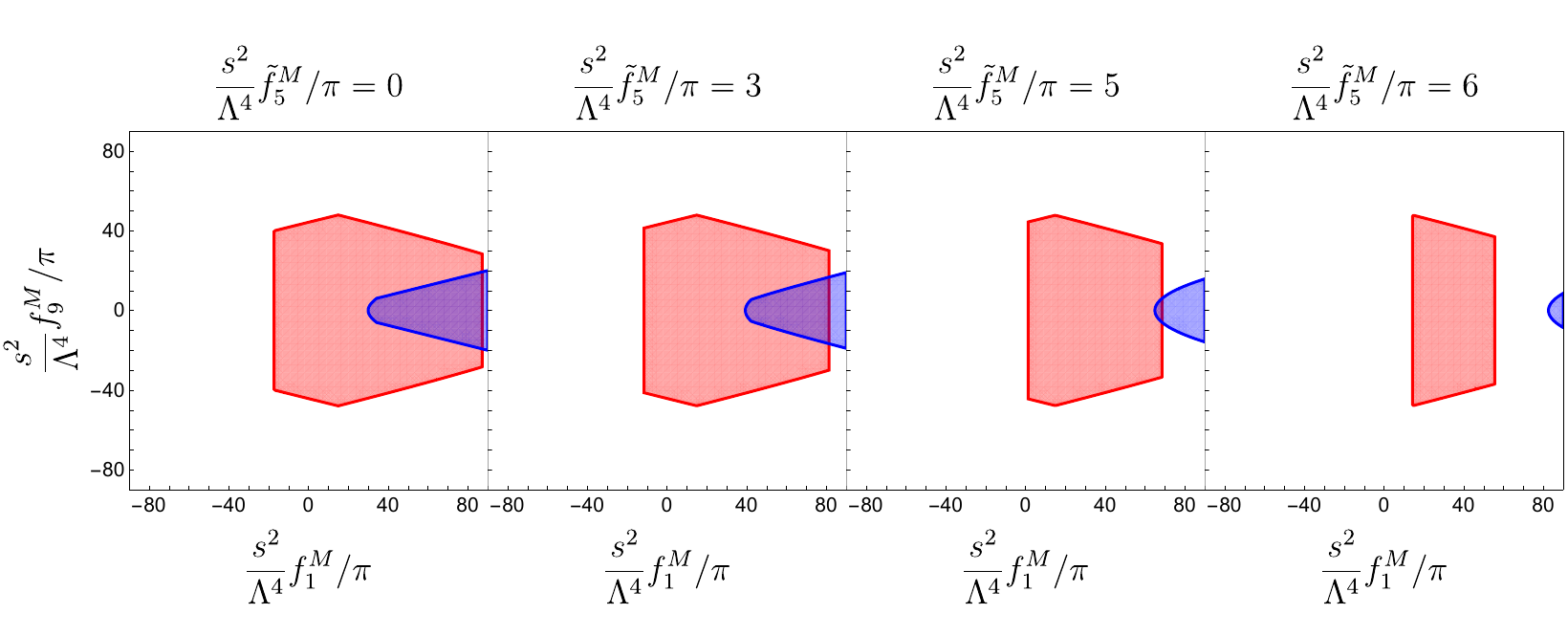}
    }

    \vspace{2mm}

    \refstepcounter{figure}
    \label{fig:positivity-unitarity-interplay}

    \makebox[\textwidth][c]{%
    \begin{minipage}{0.92\textwidth}
    \small
    \justifying
    \noindent\textbf{FIG.~\thefigure.}
    Interplay between positivity (blue regions) and perturbative
    unitarity (red regions) constraints in two-dimensional slices of
    Wilson-coefficient space. The axes show the dimensionless rescaled CP-even coefficients $s^2 f_i/(\pi\Lambda^4)$ and CP-odd coefficients $s^2\tilde{f}_i/(\pi\Lambda^4)$. The parameter space compatible with both requirements is given by the overlap; white regions are excluded. The top panels consider the $W^4$ sector and they display the allowed regions of the rescaled $(\tilde{f}^{T}_5, \tilde{f}^{T}_6)$ plane, with
    $s^2 f^{T}_2/(\pi\Lambda^4)=0,3,6,9$ from left to right panels
    while all other coefficients are set to zero. The case of the $\Phi^2W^2$ sector is shown in the bottom panels that show the limits on the rescaled $(f^{M}_1, f^{M}_9)$ plane, with
    $s^2\tilde{f}^{M}_5/(\pi\Lambda^4)=0,3,5,6$ from left to right
    panels, while $s^2 f^{M}_0/(\pi\Lambda^4)=5$, $s^2 f^{M}_7/(\pi\Lambda^4)=30$, and all other coefficients are set to zero.

  \end{minipage}%
    }

\end{figure*}

To better understand the interplay between the positivity and unitarity bounds we present in Fig.~\ref{fig:positivity-unitarity-interplay} representative two-dimensional slices of the Wilson-coefficient space associated with the $W^4$ (top) and $\Phi^2W^2$ (bottom) sectors. In this figure, the blue and red regions correspond, respectively, to the allowed positivity and unitarity domains, while their overlap defines the viable parameter space. The top panels depict the bounds on the rescaled ($\tilde{f}^T_5$, $\tilde{f}^T_6$) for several values of $f_2^T$ and assuming that the remaining Wilson coefficients of the $W^4$ sector vanish. The top left panel corresponds to a scenario where only the couplings $\tilde{f}^T_5$ and $\tilde{f}^T_6$ are non-vanishing which has less stringent unitarity bounds, however, positivity requires both CP-odd coefficients to vanish, leaving no nontrivial allowed region. It is clear from the other top panels that the unitarity allowed region shrinks as $f_2^T$ increases while the parameter region allowed by positivity gets larger.\smallskip
  
The bottom panels of Fig.~\ref{fig:positivity-unitarity-interplay} display the unitarity and positivity allowed regions in the rescaled ($f^M_1$, $f^M_9$) plane for several values of $\tilde{f}_5^M$, with $s^2 f^{M}_0/(\pi\Lambda^4)=5$, $s^2 f^{M}_7/(\pi\Lambda^4) =30$, and all other coefficients set to zero. Increasing $\tilde{f}_5^M$ reduces the allowed areas within
the displayed parameter range and their overlap, which vanishes in the bottom-right panel.

\section{Discussion}
\label{sec:discussion}

The study of quartic gauge couplings provides a direct probe of the
EW symmetry-breaking sector and of possible heavy new physics
encoded in the SMEFT. Dimension-eight genuine aQGC operators induce
amplitudes that grow as $s^2$, and therefore consistency of the EFT
description and correspondence with a possible UV completion impose
nontrivial restrictions on the Wilson-coefficient space. In this work we have extended the perturbative unitarity analysis of CP-even aQGC operators performed in Ref.~\cite{Almeida:2020ylr} to the operators that are C- or P-odd, and supplemented it with forward-limit positivity constraints. \smallskip

We have presented the results in three benchmark scenarios. The
results in Table~\ref{tab:ls_individual} correspond to the one
non-vanishing aQGC operator scenario using only perturbative
unitarity. These bounds permit a direct comparison with previous
studies of the C- and P-even sector and provide useful checks of the
single-operator assumptions commonly adopted in experimental analyses. However, they should only be understood as reference limits. In particular, CP-odd coefficients do not define generic
positivity-compatible directions by themselves as shown in
Eq.~(\ref{fff}), implying that nonzero CP-odd effects typically
require CP-even support. The comparison between Tables~\ref{tab:ls_individual}, \ref{tab:multi} and
Tables~\ref{tab:multi2}, \ref{tab:multi3} shows that positivity
restricts the allowed directions in Wilson-coefficient space and
therefore reduces the viable parameter space. \smallskip

This point is particularly important for the $\Phi^2BW$ and $B^2W^2$
sectors. From the point of view of perturbative unitarity, these
sectors can be used to organize the coupled-channel amplitudes and
obtain marginalized bounds. However, with regard to positivity they
cannot be treated as isolated sectors if one wants to allow generic
nonzero coefficients. In particular, nonzero off-diagonal
contributions from $\Phi^2BW$ can require diagonal support from
$\Phi^2B^2$ and $\Phi^2W^2$. Similarly, although $B^2W^2$ already
contains elastic $BW\to BW$ amplitudes, keeping certain coefficient combinations nonzero can require additional operators from the $B^4$ and $W^4$ sectors. Therefore, forward positivity also
imposes nontrivial correlations among different operator classes,
complementing the constraints obtained from perturbative unitarity.
\smallskip

The interplay between the two requirements is illustrated in Fig.~\ref{fig:positivity-unitarity-interplay}. The blue and red regions show, respectively, the parameter space allowed by positivity and perturbative unitarity. The nontrivial overlap between these regions demonstrates that the two consistency conditions are complementary and must be imposed simultaneously to identify the physically viable EFT parameter space. The combined implications of these constraints can be explored using the Python and Mathematica codes available in the accompanying GitHub repository~\cite{aQGCchecker}. These codes implement the tabulated unitarity bounds and positivity inequalities for user-defined values of the Wilson coefficients.\smallskip

We close the discussion with an outlook to future directions. On the
one hand, it is important to stress that the positivity bounds derived in Sec.~\ref{sim_ops} constitute necessary analytical conditions in the forward limit, rather than a complete characterization of forward-elastic positivity. Establishing the latter would require a systematic treatment of the positivity of the forward-amplitude matrices for arbitrary external-state superpositions, including more involved inequalities than those considered here. We leave such an analysis for future work, where more advanced techniques, such as the extremal-ray method~\cite{Yamashita:2020gtt, Chang:2026ztn,Zhang:2020jyn}, could prove
particularly useful. Nevertheless, our results provide a systematic
treatment of partial-wave unitarity in the C- or P-odd aQGC sector of the SMEFT, complemented by analytical positivity constraints that
further restrict the allowed Wilson-coefficient space. On the other
hand, we have not addressed the corresponding questions of positivity and perturbative unitarity in the Higgs Effective Field Theory (HEFT)~\cite{Longhitano:1980tm, Alonso:2012px, Buchalla:2013rka, Brivio:2016fzo, Alonso:2015fsp, Falkowski:2019tft, Cohen:2020xca}. In contrast to the SMEFT, the HEFT organizes operators according to their chiral power counting. In particular, positivity constraints on CP-odd operators contributing to longitudinal gauge-Higgs scattering at $\mathcal O(p^4)$ have recently been investigated in Ref.~\cite{Chakraborty:2024ciu}. However, extending the analysis to the mixed and transverse genuine aQGC structures considered here requires accounting for operators at higher chiral orders~\cite{Brivio:2016fzo, Sun:2022snw}. A systematic comparison of the corresponding positivity and perturbative unitarity constraints in both effective descriptions therefore constitutes an
interesting direction for future work.\smallskip

\section*{Acknowledgments}

Á.\ L.-O.\ and S.\ R.\ H.\ are grateful for the hospitality of the Department of Mathematical Physics at the University of São Paulo, where much of this work was carried out. Á.\ L.-O.\ and S.\ R.\ H.\ also thank Jaime Fernández Tejedor and Benjamin Grinstein for two very illuminating discussions on positivity and unitarity, and Gabriele Levati for a useful email exchange on the same topics. Á.\ L.-O.\ and S.\ R.\ H.\ further acknowledge the use of the HPC facilities at IFT UAM-CSIC for some of the calculations presented in this work. O.\ J.\ P.\ E.\ is partially supported by CNPq Grant No.\ 302120/2025-4. The work of Á.\ L.-O.\ is supported by FPI Grant No.\ PRE2022-105383, funded by MCIN/AEI/10.13039/501100011033 and ESF+. The work of S.\ R.\ H.\ is supported by FPI Grant No.\ PREP2022-000805, funded by MCIN/AEI/10.13039/501100011033. Á.\ L.-O.\ and S.\ R.\ H.\ further acknowledge support from the Spanish Research Agency (Agencia Estatal de Investigación) through the grant IFT Centro de Excelencia Severo Ochoa CEX2020-001007-S, funded by MCIN/AEI/10.13039/501100011033. This project has received funding from the European Union's Horizon Europe research and innovation programme under the Marie Skłodowska-Curie Staff Exchanges grant agreement No.\ 101086085 - ASYMMETRY.

%
\bibliographystyle{apsrev4-2}
\bibliography{quarticreferences}
%

\appendix

\clearpage
\section{Helicity amplitudes}
\label{app:tables}

Here we provide the independent helicity amplitudes $\mathcal{M}({P_1}_{\lambda_1}{P_2}_{\lambda_2}\to{P_3}_{\lambda_3}{P_4}_{\lambda_4})$ for the non-vanishing $2\to2$ scatterings induced by the C- or P-odd operators. The corresponding amplitudes for the C- and P-even operators can be found in Ref.~\cite{Almeida:2020ylr}. Unlike the amplitudes tabulated in that reference, here we present them in an unsymmetrized ordered basis, without the identical-particle normalization factors entering the partial-wave expansion of Eq.~\eqref{eq:helamp}. This convention also facilitates the construction of the forward-amplitude matrices used in the positivity analysis.

The amplitudes are organized separately for mixed and tensorial operators. The mixed-operator amplitudes are displayed in Tables~\ref{tab:ampm1} and~\ref{tab:ampm2}, while the tensorial ones are given in Table~\ref{tab:ampt1}. We use $X\equiv\cos\theta$, $X_M\equiv1-X$, and $X_P\equiv1+X$, where $\theta$ is the center-of-mass scattering angle between the momenta of particles $P_1$ and $P_3$. The azimuthal angle is set to zero.\smallskip

The amplitudes are listed for a representative set of external-particle orderings and helicity configurations. Amplitudes related by final-state permutations, or by parity and time-reversal transformations, can be reconstructed from those provided. Exchanging the two final particles, together with their momenta and helicity labels, corresponds to the interchange $t\leftrightarrow u$, or equivalently $X\to-X$ in the massless limit. Taking into account the helicity-basis phase associated with our Jacob--Wick convention~\cite{Jacob:1959at}, the exchange relation reads
\begin{equation}
    \mathcal{M}\left({P_1}_{\lambda_1}{P_2}_{\lambda_2}\to {P_3}_{\lambda_3}{P_4}_{\lambda_4}\right)(s,\theta)
    =\chi_{34}\,
    \mathcal{M}\left({P_1}_{\lambda_1}{P_2}_{\lambda_2}\to {P_4}_{\lambda_4}{P_3}_{\lambda_3}\right)(s,\pi-\theta) \,,
\end{equation}
where, in the conventions used throughout this work,
\begin{equation}
    \chi_{34}=(-1)^{j_3+j_4-\lambda_3-\lambda_4} \, ,
\end{equation}
with $j_i$ denoting the spin of the corresponding external particle.\smallskip

The parity and time-reversal relations for the helicity amplitudes are obtained from Eqs.~\eqref{parity} and~\eqref{time} in Sec.~\ref{sec:unit}, together with the transformation properties of the Wigner $d$ matrices entering the partial-wave expansion in Eq.~\eqref{eq:helamp}. These relations are applied separately to each operator contribution, taking into account its corresponding parity and time-reversal eigenvalues.\smallskip

Crossing relations could be used to further reduce the number of independent channels (cf.\ Appendix~E of Ref.~\cite{Hebbar:2020ukp}). However, for clarity, we restrict the reconstruction to final-state permutations and the parity and time-reversal relations described above.

\begingroup
  \squeezetable
  \begin{table}
\begin{tabular} {|c||c|l|}
\hline
&\multicolumn{2}{c|}{\rule{0pt}{4.5ex}$\displaystyle\times \frac{s^2}{\Lambda^4}$}\\[2mm]\hline

$W^+W^-\rightarrow W^+W^-$    
& $-\!-0\,\,0$ & \rule{0pt}{3ex}$\frac{i}{4}(4\tilde{f}^M_5-X\tilde{f}^M_6)$\\
& $-\,0-0$ & \rule{0pt}{3ex}$-\frac{1}{4}X_P f^M_9$\\
& $-\,0+0$ & \rule{0pt}{3ex}$-\frac{i}{16}X_M((X_M-4)\tilde{f}^M_6-4X_M\tilde{f}^M_5)$\\
& $-\!+0\,\,0$ & \rule{0pt}{3ex}$\frac{1}{8}(1-X^2)f^M_9$\\
& $0-0\,-$ & \rule{0pt}{3ex}$\frac{1}{4}X_P f^M_9$\\
& $0-0\,+$ & \rule{0pt}{3ex}$-\frac{i}{16}X_M((X_M-4)\tilde{f}^M_6-4X_M\tilde{f}^M_5)$\\[1mm]
\hline

$W^+W^-\rightarrow ZZ$
& $-\!-0\,\,0$ & \rule{0pt}{3ex}$i\tilde{f}^M_5$\\
& $-\,0-0$ & \rule{0pt}{3ex}$-\frac{1}{8}X_P(2c_W f^M_9+i s_W\tilde{f}^M_{4})$\\
& $-\,0\,\,0\,+$ & \rule{0pt}{3ex}$-\frac{1}{32}X_P((X_P-4)(s_W(f^M_8-i\tilde{f}^M_3)-2i c_W \tilde{f}^M_6)+i X_P s_W(2\tilde{f}^M_2+\tilde{f}^M_4))$\\
& $0\,\,0-\!-$ & \rule{0pt}{3ex}$-i(s_W^2\tilde{f}^M_1-c_W^2\tilde{f}^M_5+\frac{1}{8}s_{2W}(2\tilde{f}^M_2+\tilde{f}^M_4))$\\[1mm]
\hline

$W^+W^-\rightarrow Z\gamma$
& $-\,0\,\,0\,-$ & \rule{0pt}{3ex}$\frac{1}{8}X_M(2 s_W f^M_9-i c_W\tilde{f}^M_{4})$\\
& $-\,0\,\,0\,+$ & \rule{0pt}{3ex}$\frac{1}{32}X_P((X_P-4)(c_W(f^M_8-i\tilde{f}^M_3)+2i s_W \tilde{f}^M_6)+i X_P c_W(2\tilde{f}^M_2+\tilde{f}^M_4))$\\
& $0-0\,-$ & \rule{0pt}{3ex}$\frac{1}{8}X_P (2s_W f^M_9+i c_W\tilde{f}^M_{4})$\\
& $0-0\,+$ & \rule{0pt}{3ex}$\frac{1}{32}X_M((X_M-4)(c_W(f^M_8-i\tilde{f}^M_3)-2i s_W \tilde{f}^M_6)-i X_M c_W(2\tilde{f}^M_2+\tilde{f}^M_4))$\\
& $0\,\,0-\!-$ & \rule{0pt}{3ex}$\frac{1}{8}(X(f^M_8+i\tilde{f}^M_3)+i(4 s_{2W}(\tilde{f}^M_1-\tilde{f}^M_5)-c_{2W}(2\tilde{f}^M_2+\tilde{f}^M_4)))$\\
& $0\,\,0-\!+$ & \rule{0pt}{3ex}$-\frac{i}{16}(1-X^2)\tilde{f}^M_4$\\[1mm]
\hline

$W^+W^-\rightarrow ZH$
& $-\!-0\,\,0$ & \rule{0pt}{3ex}$-\frac{i}{4}X\tilde{f}^M_{6}$\\
& $-\,0-0$ & \rule{0pt}{3ex}$\frac{1}{8} X_P(2 c_W f^M_9+i s_W\tilde{f}^M_4)$\\
& $-\,0+0$ & \rule{0pt}{3ex}$-\frac{1}{32}X_M((X_M-4)(s_W(f^M_8-i\tilde{f}^M_3)-2i c_W\tilde{f}^M_6)+i X_M s_W(2\tilde{f}^M_2+\tilde{f}^M_4))$\\
& $0-\!-\,0$ & \rule{0pt}{3ex}$-\frac{1}{8} X_M(2 c_W f^M_9-i s_W \tilde{f}^M_4)$\\
& $0-\!+\,0$ & \rule{0pt}{3ex}$\frac{1}{32}X_P((X_P-4)(s_W(f^M_8-i\tilde{f}^M_3)+2i c_W\tilde{f}^M_6)-i X_P s_W(2\tilde{f}^M_2+\tilde{f}^M_4))$\\[1mm]
\hline

$W^+W^-\rightarrow \gamma\gamma$
& $0\,\,0-\!-$ & \rule{0pt}{3ex}$-i(c_W^2\tilde{f}^M_1+s_W^2\tilde{f}^M_5+\frac{1}{8}s_{2W}(2\tilde{f}^M_2+\tilde{f}^M_4))$\\[1mm]
\hline

$W^+W^-\rightarrow \gamma H$ 
& $-\,0-0$ & \rule{0pt}{3ex}$\frac{1}{8}X_P(2 s_W f^M_9-i c_W\tilde{f}^M_{4})$\\
& $-\,0+0$ & \rule{0pt}{3ex}$\frac{1}{32} X_M((X_M-4)(c_W(f^M_8-i\tilde{f}^M_3)+2i s_W\tilde{f}^M_6)+i X_M c_W(2\tilde{f}^M_2+\tilde{f}^M_4))$\\
& $0-\!-\,0$ & \rule{0pt}{3ex}$-\frac{1}{8}X_M (2 s_W f^M_9+i c_W\tilde{f}^M_{4})$\\
& $0-\!+\,0$ & \rule{0pt}{3ex}$-\frac{1}{32} X_P((X_P-4)(c_W(f^M_8-i\tilde{f}^M_3)-2i s_W\tilde{f}^M_6)-i X_P c_W(2\tilde{f}^M_2+\tilde{f}^M_4))$\\[1mm]
\hline

$W^+W^-\rightarrow HH$ 
& $-\!-0\,\,0$ & \rule{0pt}{3ex}$-i\tilde{f}^M_5$\\
& $-\!+0\,\,0$ & \rule{0pt}{3ex}$\frac{1}{8}(1-X^2)f^M_9$\\[1mm]
\hline

$ZZ\rightarrow ZZ$
& $-\!-0\,\,0$ & \rule{0pt}{3ex} $i(c_W^2\tilde{f}^M_{5} +s_W^2\tilde{f}^M_1+ \frac{1}{8}s_{2W}(2\tilde{f}^M_{2} + \tilde{f}^M_{4}))$\\
& $-\,0\,\,0\,+$ & \rule{0pt}{3ex} $-\frac{i}{4}X_P^2(c_W^2\tilde{f}^M_{5} +s_W^2\tilde{f}^M_1+ \frac{1}{8}s_{2W}(2\tilde{f}^M_{2} + \tilde{f}^M_{4}))$\\[1mm]
\hline

$ZZ\rightarrow Z\gamma$
& $-\,0\,\,0\,-$ & \rule{0pt}{3ex} $-\frac{i}{8}X_M\tilde{f}^M_4$\\
& $-\,0\,\,0\,+$ & \rule{0pt}{3ex} $\frac{i}{32}X_P^2(c_{2W}(2\tilde{f}^M_2+\tilde{f}^M_4)+4s_{2W}(\tilde{f}^M_1-\tilde{f}^M_5))$\\
& $0\,\,0-\!-$ & \rule{0pt}{3ex} $\frac{i}{8}(c_{2W}(2\tilde{f}^M_2+\tilde{f}^M_4)+4s_{2W}(\tilde{f}^M_1-\tilde{f}^M_5))$\\
& $0\,\,0-\!+$ & \rule{0pt}{3ex} $\frac{i}{16}(1-X^2)\tilde{f}^M_4$\\[1mm]
\hline

$ZZ\rightarrow \gamma\gamma$ 
& $0\,\,0-\!-$ &  \rule{0pt}{3ex} $-i(c_W^2\tilde{f}^M_1+s_W^2\tilde{f}^M_{5} -\frac{1}{8}s_{2W}(2\tilde{f}^M_{2} + \tilde{f}^M_{4}))$\\[1mm]
\hline

$ZZ\rightarrow \gamma H$ 
& $-\,0+0$ &  \rule{0pt}{3ex} $\frac{1}{32}(X_P^2-4)(f^M_8-i \tilde{f}^M_3)$\\[1mm]
\hline

$ZZ\rightarrow HH$
& $-\!-0\,\,0$ & \rule{0pt}{3ex} $-i(s_W^2\tilde{f}^M_1+c_W^2\tilde{f}^M_{5}+\frac{1}{8}s_{2W}(2\tilde{f}^M_{2} + \tilde{f}^M_{4}))$\\[1mm]
\hline

$Z\gamma\rightarrow Z\gamma$ 
& $0-0\,+$ & \rule{0pt}{3ex} $\frac{i}{4}X_M^2(c_W^2\tilde{f}^M_1+s_W^2\tilde{f}^M_{5}-\frac{1}{8}s_{2W}(2\tilde{f}^M_{2} + \tilde{f}^M_{4}))$\\[1mm]
\hline

$Z\gamma\rightarrow ZH$ 
& $-\!-0\,\,0$ & \rule{0pt}{3ex} $-\frac{1}{8}X(f^M_8-i \tilde{f}^M_3)$\\
& $0-\!+\,0$ & \rule{0pt}{3ex} $\frac{1}{32}(X_M^2-4)(f^M_8-i \tilde{f}^M_3)$\\[1mm]
\hline

$Z\gamma\rightarrow HH$ 
& $-\!-0\,\,0$ & \rule{0pt}{3ex} $\frac{i}{8}(c_{2W}(2\tilde{f}^M_{2} + \tilde{f}^M_{4})+4s_{2W}(\tilde{f}^M_1-\tilde{f}^M_{5}))$\\
& $-\!+0\,\,0$ & \rule{0pt}{3ex} $\frac{i}{16}(1-X^2)\tilde{f}^M_4$\\[1mm]
\hline

$ZH\rightarrow ZH$ 
& $-\,0+0$ & \rule{0pt}{3ex} $\frac{i}{4}X_M^2(s_W^2\tilde{f}^M_1+c_W^2\tilde{f}^M_{5} +\frac{1}{8}s_{2W}(2\tilde{f}^M_{2} + \tilde{f}^M_{4}))$\\[1mm]
\hline

$ZH\rightarrow \gamma H$ 
& $-\,0-0$ & \rule{0pt}{3ex} $\frac{i}{8}X_P\tilde{f}^M_4$\\
& $-\,0+0$ & \rule{0pt}{3ex} $-\frac{i}{32}X_M^2(c_{2W}(2\tilde{f}^M_{2} + \tilde{f}^M_{4})+4s_{2W}(\tilde{f}^M_1-\tilde{f}^M_{5}))$\\[1mm]
\hline

$\gamma\gamma\rightarrow H H$
& $-\!-0\,\,0$ & \rule{0pt}{3ex} $-i(c_W^2\tilde{f}^M_1+s_W^2\tilde{f}^M_{5}-\frac{1}{8}s_{2W}(2\tilde{f}^M_{2} + \tilde{f}^M_{4}))$\\[1mm]
\hline

$\gamma H\rightarrow \gamma H$ 
& $-\,0+0$ & \rule{0pt}{3ex} $\frac{i}{4}X_M^2(c_W^2\tilde{f}^M_1+s_W^2\tilde{f}^M_{5}-\frac{1}{8}s_{2W}(2\tilde{f}^M_{2} + \tilde{f}^M_{4}))$\\[1mm]
\hline
\end{tabular}

\refstepcounter{table}

\vspace{2mm}

\makebox[\textwidth][c]{%
\begin{minipage}{0.95\textwidth}
\small
\justifying
\noindent\textbf{TABLE~\thetable.}
Unitarity-violating (growing as $s^2$) terms of the scattering amplitudes $\mathcal{M}({P_1}_{\lambda_1}{P_2}_{\lambda_2}\to{P_3}_{\lambda_3}{P_4}_{\lambda_4})$ for the gauge boson helicities given in the second column generated by the mixed operators that are C- or P-odd presented in Sec.~\ref{sec:oper}. Here $X\equiv\cos\theta$, $X_M\equiv1-\cos\theta$, $X_P\equiv1+\cos\theta$, $c_W=\cos\theta_W$, $s_W=\sin\theta_W$, $c_{2W}=\cos2\theta_W$, $s_{2W}=\sin2\theta_{W}$, and the overall factor extracted from all amplitudes is given at the top of the table.
\end{minipage}%
}
\label{tab:ampm1}
\end{table}
\endgroup

\begingroup
  \squeezetable
  \begin{table}
\begin{tabular} {|c||c|l|}
\hline
&\multicolumn{2}{c|}{\rule{0pt}{4.5ex}$\displaystyle\times \frac{s^2}{\Lambda^4}$}\\[2mm]\hline
$W^+W^+\rightarrow W^+W^+$
& $-\,0-0$ & \rule{0pt}{3ex}$-\frac{1}{8}X_P f^M_9$\\[1mm]
& $-\,0\,\,0\,+$ & \rule{0pt}{3ex}$-\frac{i}{16}X_P\left(4X_P\tilde{f}^M_5+\left(X_P-4\right)\tilde{f}^M_6\right)$\\[1mm]
\hline

$W^+Z\rightarrow W^+Z$ 
& $-\!-0\,\,0$ & \rule{0pt}{3ex}$\frac{1}{8}\left(X\left(s_W\left(f^M_8-i\tilde{f}^M_3\right)-2i c_W\tilde{f}^M_6\right)+i s_W\left(2\tilde{f}^M_2+\tilde{f}^M_4\right)\right)$\\[1mm]
& $-\,0-0$ & \rule{0pt}{3ex}$\frac{1}{4}X_P f^M_9$\\[1mm]
& $-\,0\,\,0\,-$ & \rule{0pt}{3ex}$\frac{1}{8}X_M\left(2c_W f^M_9+i s_W\tilde{f}^M_4\right)$\\[1mm]
& $-\,0\,\,0\,+$ & \rule{0pt}{3ex}$\frac{1}{32}X_P\left(\left(X_P-4\right)\left(s_W\left(f^M_8-i\tilde{f}^M_3\right)-2i c_W\tilde{f}^M_6\right)-i X_P s_W\left(2\tilde{f}^M_2+\tilde{f}^M_4\right)\right)$\\[1mm]
& $-\,0+0$ & \rule{0pt}{3ex}$\frac{i}{4} X_M^2\tilde{f}^M_5$\\[1mm]
& $-\!+0\,\,0$ & \rule{0pt}{3ex}$\frac{1}{16}\left(1-X^2\right)\left(2c_W f^M_9+i s_W\tilde{f}^M_4\right)$\\[1mm]
& $0-0\,+$ & \rule{0pt}{3ex}$\frac{i}{4}X_M^2\left(s_W^2\tilde{f}^M_1+c_W^2\tilde{f}^M_5-\frac{1}{8}s_{2W}\left(2\tilde{f}^M_2+\tilde{f}^M_4\right)\right)$ \\[1mm]
\hline

$W^+Z\rightarrow W^+\gamma$
& $-\,0\,\,0\,-$ & \rule{0pt}{3ex}$\frac{1}{8}X_M\left(2s_W f^M_9-i c_W\tilde{f}^M_4\right)$\\[1mm]
& $-\,0\,\,0\,+$ & \rule{0pt}{3ex}$-\frac{1}{32}X_P\left(\left(X_P-4\right)\left(c_W\left(f^M_8-i\tilde{f}^M_3\right)+2i s_W\tilde{f}^M_6\right)-i X_P c_W\left(2\tilde{f}^M_2+\tilde{f}^M_4\right)\right)$\\[1mm]
& $0-0\,-$ & \rule{0pt}{3ex}$-\frac{i}{8}X_P\tilde{f}^M_4$\\[1mm]
& $0-0\,+$ & \rule{0pt}{3ex}$-\frac{1}{32}X_M\left(\left(X_M-4\right)\left(f^M_8-i\tilde{f}^M_3\right)+i X_M\left(4s_{2W}\left(\tilde{f}^M_1-\tilde{f}^M_5\right)-c_{2W}\left(2\tilde{f}^M_2+\tilde{f}^M_4\right)\right)\right)$\\[1mm]
& $0\,\,0-\!-$ & \rule{0pt}{3ex}$-\frac{1}{8}\left(X\left(c_W\left(f^M_8+i\tilde{f}^M_3\right)-2i s_W\tilde{f}^M_6\right)-i c_W\left(2\tilde{f}^M_2+\tilde{f}^M_4\right)\right)$\\[1mm]
& $0\,\,0-\!+$ & \rule{0pt}{3ex}$\frac{1}{16}\left(1-X^2\right)\left(2s_W f^M_9+i c_W\tilde{f}^M_4\right)$\\[1mm]
\hline

$W^+Z\rightarrow W^+H$
& $-\!-0\,\,0$ & \rule{0pt}{3ex}$-\frac{1}{8}\left(X\left(s_W\left(f^M_8-i\tilde{f}^M_3\right)-2i c_W\tilde{f}^M_6\right)+is_W\left(2\tilde{f}^M_2+\tilde{f}^M_4\right)\right)$\\[1mm]
& $-\,0+0$ & \rule{0pt}{3ex}$-\frac{i}{16}\left(X_P^2-4\right)\tilde{f}^M_6$\\[1mm]
& $-\!+0\,\,0$ & \rule{0pt}{3ex}$-\frac{1}{16}\left(1-X^2\right)\left(2c_W f^M_9+i s_W\tilde{f}^M_4\right)$\\[1mm]
& $0-\!-\,0$ & \rule{0pt}{3ex}$\frac{1}{8}X_M \left(2c_W f^M_9-i s_W\tilde{f}^M_4\right)$\\[1mm]
& $0-\!+\,0$ & \rule{0pt}{3ex}$-\frac{1}{32}X_P\left(\left(X_P-4\right)\left(s_W\left(f^M_8-i\tilde{f}^M_3\right)+2i c_W\tilde{f}^M_6\right)+i X_P s_W\left(2\tilde{f}^M_2+\tilde{f}^M_4\right)\right)$\\[1mm]
\hline

$W^+\gamma\rightarrow W^+\gamma$
& $0-0\,+$ & \rule{0pt}{3ex}$\frac{i}{4}X_M^2\left(c_W^2\tilde{f}^M_1+s_W^2\tilde{f}^M_5+\frac{1}{8}s_{2W}\left(2\tilde{f}^M_2+\tilde{f}^M_4\right)\right)$\\[1mm]
\hline

$W^+\gamma\rightarrow W^+H$
& $-\!-0\,\,0$ & \rule{0pt}{3ex}$\frac{1}{8}\left(X\left(c_W\left(f^M_8-i\tilde{f}^M_3\right)+2i s_W\tilde{f}^M_6\right)+i c_W\left(2\tilde{f}^M_2+\tilde{f}^M_4\right)\right)$\\[1mm]
& $-\!+0\,\,0$ & \rule{0pt}{3ex}$-\frac{1}{16}\left(1-X^2\right)\left(2s_W f^M_9-i c_W\tilde{f}^M_4\right)$\\[1mm]
& $0-\!-\,0$ & \rule{0pt}{3ex}$\frac{1}{8}X_M \left(2s_W f^M_9+i c_W\tilde{f}^M_4\right)$\\[1mm]
& $0-\!+\,0$ & \rule{0pt}{3ex}$\frac{1}{32}X_P\left(\left(X_P-4\right)\left(c_W\left(f^M_8-i\tilde{f}^M_3\right)-2i s_W\tilde{f}^M_6\right)+i X_P c_W\left(2\tilde{f}^M_2+\tilde{f}^M_4\right)\right)$\\[1mm]
\hline

$W^+H\rightarrow W^+H$
& $-\,0-0$ & \rule{0pt}{3ex}$\frac{1}{4}X_P f^M_9$\\[1mm]
& $-\,0+0$ & \rule{0pt}{3ex}$\frac{i}{4}X_M^2\tilde{f}^M_5$\\[1mm]
\hline

\end{tabular}

\refstepcounter{table}

\vspace{2mm}

\makebox[\textwidth][c]{%
\begin{minipage}{0.95\textwidth}
\small
\justifying
\noindent\textbf{TABLE~\thetable.}
Continuation of Table~\ref{tab:ampm1} using the same conventions.
\end{minipage}%
}
\label{tab:ampm2}
  \end{table}
  \endgroup  

  \begingroup
  \squeezetable
  \begin{table}
\begin{tabular} {|c||c|l|}
\hline
&\multicolumn{2}{c|}{\rule{0pt}{4.5ex}$\displaystyle\times i \frac{s^2}{\Lambda^4}$}\\[2mm]\hline

$W^+W^+ \rightarrow W^+W^+$
&$--+\,+$ & \rule{0pt}{3ex}$4\left(1+X^2\right)\tilde{f}^T_{5}+2\left(5+X^2\right)\tilde{f}^T_{6}$\\[1mm]
\hline

$W^+Z \rightarrow W^+Z$
&$--+\,+$ & \rule{0pt}{3ex}$X_M^2\left(s_W^2\left(\tilde{f}^T_2+\tilde{f}^T_3\right)+2c_W^2\tilde{f}^T_5\right)+\frac{1}{2}\left(4+X_P^2\right)\left(s_W^2\tilde{f}^T_4+2c_W^2\tilde{f}^T_6\right)$\\[1mm]
&$-++\,-$& \rule{0pt}{3ex}$-X_M^2 s_W^2\left(\tilde{f}^T_{2}-\tilde{f}^T_{3}\right)$ \\[1mm]
\hline
$W^+Z \rightarrow W^+\gamma$
&$--+\,+$& \rule{0pt}{3ex}$-\frac{1}{4}s_{2W}\left(2X_M^2\left(\tilde{f}^T_2+\tilde{f}^T_3-2\tilde{f}^T_5\right)+\left(4+X_P^2\right)\left(\tilde{f}^T_4-2\tilde{f}^T_6\right)\right)$\\[1mm]
&$-++\,-$& \rule{0pt}{3ex}$\frac{1}{2}X_M^2 s_{2W}\left(\tilde{f}^T_2-\tilde{f}^T_3\right)$\\[1mm]
\hline
$W^+\gamma \rightarrow W^+\gamma$ 
&$--+\,+$ & \rule{0pt}{3ex}$X_M^2\left(c_W^2\left(\tilde{f}^T_2+\tilde{f}^T_3\right)+2s_W^2\tilde{f}^T_5\right)+\frac{1}{2}\left(4+X_P^2\right)\left(c_W^2\tilde{f}^T_4+2s_W^2\tilde{f}^T_6\right)$\\[1mm]
&$-++\,-$& \rule{0pt}{3ex}$-X_M^2 c_W^2\left(\tilde{f}^T_{2}-\tilde{f}^T_{3}\right)$ \\[1mm]
\hline

$W^+W^- \rightarrow W^+W^-$
&$--+\,+$& \rule{0pt}{3ex}$2\left(4+X_M^2\right)\tilde{f}^T_5+\left(4X_M+3X_P^2\right)\tilde{f}^T_6$\\[1mm]
\hline

$W^+W^- \rightarrow ZZ$ 
&$---\,-$& \rule{0pt}{3ex}$-4s_W^2\left(\tilde{f}^T_{2}-\tilde{f}^T_{3}\right)$ \\[1mm]
&$--+\,+$& \rule{0pt}{3ex}$4 \left(s_W^2\left(\tilde{f}^T_2+\tilde{f}^T_3\right)+2 c_W^2\tilde{f}^T_5+\frac{1}{4}\left(1+X^2\right)\left(s_W^2\tilde{f}^T_4+2c_W^2\tilde{f}^T_6\right)\right)$ \\[1mm]
\hline

$W^+W^- \rightarrow Z\gamma$
&$---\,-$& \rule{0pt}{3ex}$2 s_{2W}\left(\tilde{f}^T_2-\tilde{f}^T_3\right)$\\[1mm]
&$--+\,+$& \rule{0pt}{3ex}$-\frac{1}{2}s_{2W}\left(4\tilde{f}^T_2+4\tilde{f}^T_3-8\tilde{f}^T_5+\left(1+X^2\right)\left(\tilde{f}^T_4- 2\tilde{f}^T_6\right)\right)$\\[1mm]
\hline

$W^+W^- \rightarrow \gamma\gamma$
&$---\,-$& \rule{0pt}{3ex}$-4c_W^2\left(\tilde{f}^T_{2}-\tilde{f}^T_{3}\right)$ \\[1mm]
&$--+\,+$& \rule{0pt}{3ex}$4 \left(c_W^2\left(\tilde{f}^T_2+\tilde{f}^T_3\right)+2 s_W^2\tilde{f}^T_5+\frac{1}{4}\left(1+X^2\right)\left(c_W^2\tilde{f}^T_4+2s_W^2\tilde{f}^T_6\right)\right)$ \\[1mm]
\hline     

$ZZ \rightarrow ZZ$
&$--+\,+$& \rule{0pt}{3ex}
    $\left(3+X^2\right) \left(s_{2W}^2\left(\tilde{f}^T_{2}+\tilde{f}^T_{3}+ \tilde{f}^T_{4}\right) + 4c_W^4 \left(\tilde{f}^T_{5}+\tilde{f}^T_{6}\right)+ 4s_W^4\tilde{f}^T_{1}\right)$\\[1mm]
\hline    

$ZZ \rightarrow Z\gamma$
&$---\,-$& \rule{0pt}{3ex} 
    $2 s_{2W} \left(\tilde{f}^T_{2}-\tilde{f}^T_{3}\right)$ \\[1mm]
&$--+\,+$& \rule{0pt}{3ex}
    $\left(3+X^2\right) s_{2W} \left(\tilde{f}^{T}_{2}+\tilde{f}^{T}_{3}+\tilde{f}^{T}_{4}-2 s_W^2\tilde{f}^{T}_{1}-2c_W^2\left(\tilde{f}^{T}_{2}+\tilde{f}^{T}_{3}+\tilde{f}^{T}_{4}-\tilde{f}^{T}_{5}-\tilde{f}^{T}_{6}\right)\right)$\\[1mm]
&$-+-\,+$& \rule{0pt}{3ex} $-\frac{1}{2}X_P^2 s_{2W} \left(\tilde{f}^T_{2}-\tilde{f}^T_{3}\right)$\\[1mm]
&$-++\,-$& \rule{0pt}{3ex} $\frac{1}{2}X_M^2 s_{2W} \left(\tilde{f}^T_{2}-\tilde{f}^T_{3}\right)$\\[1mm]
\hline   

$ZZ \rightarrow \gamma\gamma$
&$---\,-$& \rule{0pt}{3ex} $-4c_{2W}\left(\tilde{f}^T_2-\tilde{f}^T_3\right)$\\[1mm]
&$--+\,+$& \rule{0pt}{3ex} $4\left(c_W^4+s_W^4\right)\left(\tilde f^{T}_{2}+\tilde f^{T}_{3}\right)+2s_{2W}^2\left(\tilde f^{T}_{1}-\tilde f^{T}_{4}+\tilde f^{T}_{5}+\tilde f^{T}_{6}\right)+$\\[1mm]
&& \rule{0pt}{3ex} $\quad+\left(1+X^2\right)\left(c_{2W}^2\tilde f^{T}_{4}+s_{2W}^2\left(\tilde f^{T}_{1}-\tilde f^{T}_{2}-\tilde f^{T}_{3}+\tilde f^{T}_{5}+\tilde f^{T}_{6}\right)\right)$\\[1mm]
\hline

$Z\gamma \rightarrow Z\gamma$
&$--+\,+$& \rule{0pt}{3ex} $X_M^2\left(\left(c_W^4+s_W^4\right)\left(\tilde f^{T}_{2}+\tilde f^{T}_{3}\right)+\frac{1}{2}s_{2W}^2\left(\tilde f^{T}_{1}-\tilde f^{T}_{4}+\tilde f^{T}_{5}+\tilde f^{T}_{6}\right)\right)+$\\[1mm]
&& \rule{0pt}{3ex} $\quad+\frac{1}{2}\left(4+X_P^2\right)\left(c_{2W}^2\tilde f^{T}_{4}+s_{2W}^2\left(\tilde f^{T}_{1}-\tilde f^{T}_{2}-\tilde f^{T}_{3}+\tilde f^{T}_{5}+\tilde f^{T}_{6}\right)\right)$\\[1mm]
&$-++\,-$& \rule{0pt}{3ex} $-X_M^2 c_{2W}\left(\tilde{f}^T_2-\tilde{f}^T_3\right)$\\[1mm]
\hline

$Z\gamma \rightarrow \gamma Z$
&$--+\,+$& \rule{0pt}{3ex} $X_P^2\left(\left(c_W^4+s_W^4\right)\left(\tilde f^{T}_{2}+\tilde f^{T}_{3}\right)+\frac{1}{2}s_{2W}^2\left(\tilde f^{T}_{1}-\tilde f^{T}_{4}+\tilde f^{T}_{5}+\tilde f^{T}_{6}\right)\right)+$\\[1mm]
&& \rule{0pt}{3ex} $\quad+\frac{1}{2}\left(4+X_M^2\right)\left(c_{2W}^2\tilde f^{T}_{4}+s_{2W}^2\left(\tilde f^{T}_{1}-\tilde f^{T}_{2}-\tilde f^{T}_{3}+\tilde f^{T}_{5}+\tilde f^{T}_{6}\right)\right)$\\[1mm]
&$-+-\,+$& \rule{0pt}{3ex} $-X_P^2 c_{2W}\left(\tilde{f}^T_2-\tilde{f}^T_3\right)$\\[1mm]
\hline

$Z\gamma \rightarrow \gamma\gamma$ 
&$---\,-$& \rule{0pt}{3ex} $-2s_{2W}\left(\tilde{f}^T_2-\tilde{f}^T_3\right)$\\[1mm]
&$--+\,+$& \rule{0pt}{3ex}
    $\left(3+X^2\right) s_{2W} \left(\tilde{f}^{T}_{2}+\tilde{f}^{T}_{3}+\tilde{f}^{T}_{4}-2c_W^2\tilde{f}^T_1-2 s_W^2\left(\tilde{f}^{T}_{2}+\tilde{f}^{T}_{3}+\tilde{f}^{T}_{4}-\tilde{f}^{T}_{5}-\tilde{f}^T_6\right)\right)$\\[1mm]
&$-+-\,+$& \rule{0pt}{3ex} $-\frac{1}{2}X_P^2 s_{2W} \left(\tilde{f}^T_{2}-\tilde{f}^T_{3}\right)$\\[1mm]
&$-++\,-$& \rule{0pt}{3ex} $-\frac{1}{2}X_M^2 s_{2W} \left(\tilde{f}^T_{2}-\tilde{f}^T_{3}\right)$\\[1mm]
\hline   

$\gamma\gamma \rightarrow \gamma\gamma$
&$--+\,+$& \rule{0pt}{3ex}
    $\left(3+X^2\right) \left(s_{2W}^2 \left(\tilde{f}^T_{2}+\tilde{f}^T_{3}+ \tilde{f}^T_{4}\right) + 4s_W^4 \left(\tilde{f}^T_{5}+\tilde{f}^T_{6}\right)+ 4c_W^4\tilde{f}^T_{1}\right)$\\[1mm]
\hline
\end{tabular}

\refstepcounter{table}

\vspace{2mm}

\makebox[\textwidth][c]{%
\begin{minipage}{0.95\textwidth}
\small
\justifying
\noindent\textbf{TABLE~\thetable.}
Same as Table~\ref{tab:ampm1} but for the tensorial operators.
\end{minipage}%
}
\label{tab:ampt1}
  \end{table}
\endgroup
\clearpage
\section{Representative Directions Used in the Positivity Scan}
\label{app:uchoices}

In this appendix we summarize the representative one-particle directions used to construct the matrices $K(u)$ entering the positivity analysis. These finite choices provide necessary forward-elastic positivity conditions, without attempting to exhaust all possible external-state superpositions. In particular, complex directions are included to probe correlations involving the CP-odd operators. The construction follows the general strategy of testing positivity on selected polarization and superposition states laid out in Refs.~\cite{Zhang:2018shp,Yamashita:2020gtt}.

\subsection{Gauge-boson directions}

We first define the finite set of helicity choices used for gauge-boson directions,
\begin{equation}
    \mathcal H_1
    =
    \left\{
    (0,0),\,
    (\pi,0),\,
    \left(\frac{\pi}{2},0\right),\,
    \left(\frac{\pi}{2},\pi\right),\,
    \left(\frac{\pi}{2},\frac{\pi}{2}\right),\,
    \left(\frac{\pi}{2},-\frac{\pi}{2}\right)
    \right\} \, .
\label{eq:helicity_set_app}
\end{equation}
For the hypercharge gauge boson, the corresponding directions are
\begin{equation}
    |B(\beta,\xi)\rangle
    =
    \cos\frac{\beta}{2}\,|B_-\rangle
    +
    e^{i\xi}\sin\frac{\beta}{2}\,|B_+\rangle \, ,
    \qquad
    (\beta,\xi)\in\mathcal H_1 \, .
\label{eq:B_helicity_direction_app}
\end{equation}
For the non-abelian gauge bosons, we also allow coherent weak-isospin superpositions,
\begin{equation}
    |W(n;\beta,\xi)\rangle
    =
    \sum_{I=1}^{3}
    n_I
    \left(
        \cos\frac{\beta}{2}\,|W^I_-\rangle
        +
        e^{i\xi}\sin\frac{\beta}{2}\,|W^I_+\rangle
    \right),
    \qquad
    n=(n_1,n_2,n_3),
    \qquad
    \sum_{I=1}^{3}|n_I|^2=1,
    \qquad
    (\beta,\xi)\in\mathcal H_1.
\label{eq:Wn_direction_app}
\end{equation}
The pair $(\beta,\xi)$ fixes the helicity direction, while $n$ fixes the direction in weak-isospin space. The choices in $\mathcal H_1$, for $V=B$ or $W^I$, correspond to
\begin{equation}
\begin{array}{c|c|c|c}
\text{choice} & \beta & \xi & \text{helicity direction} \\
\hline\rule{0pt}{2.5ex}
- & 0 & 0 & |V_-\rangle \\[1mm]
+ & \pi & 0 & |V_+\rangle \\[1mm]
{\Re,+} & \pi/2 & 0 &
\left(|V_-\rangle+|V_+\rangle\right)/\sqrt2 \\[1mm]
{\Re,-} & \pi/2 & \pi &
\left(|V_-\rangle-|V_+\rangle\right)/\sqrt2 \\[1mm]
{\Im,+} & \pi/2 & \pi/2 &
\left(|V_-\rangle+i|V_+\rangle\right)/\sqrt2 \\[1mm]
{\Im,-} & \pi/2 & -\pi/2 &
\left(|V_-\rangle-i|V_+\rangle\right)/\sqrt2\\[1mm]
\end{array}
\label{eq:helicity_choices_table_app}
\end{equation}
For the weak-isospin directions, the elementary vectors are defined as
\begin{equation}
    \hat n_1=(1,0,0) \, ,
    \qquad
    \hat n_2=(0,1,0) \, ,
    \qquad
    \hat n_3=(0,0,1) \, .
\end{equation}
The set of representative $n$ vectors used in the scan is defined as $\mathcal{N}_W$ and contains the following representative directions:
\begin{equation}
\begin{array}{ll}
\text{Pure:}
&
\begin{gathered}
    \hat n_j \, ,
    \qquad
    j=1,2,3 \,.
\end{gathered}
\\[3mm]
\text{Real two-component:}
&
\begin{gathered}
    n_{jk}^{{\Re},\pm}
    =
    \dfrac{1}{\sqrt2}
    \left(\hat n_j\pm\hat n_k\right) \, ,
    \qquad
    1\le j<k\le 3 \, .
\end{gathered}
\\[3mm]
\text{Complex two-component:}
&
\begin{gathered}
    n_{jk}^{{\Im},\pm}
    =
    \dfrac{1}{\sqrt2}
    \left(\hat n_j\pm i\hat n_k\right) \, ,
    \qquad
    1\le j<k\le 3 \, .
\end{gathered}
\\[3mm]
\text{Fully coherent:}
&
\begin{gathered}
    n_{\rm all}^{s_2s_3}
    =
    \dfrac{1}{\sqrt3}
    \left(1,s_2,s_3\right),\qquad
    n_{\rm all}^{\rm ph,\,s_2s_3}
    =
    \dfrac{1}{\sqrt3}
    \left(1,s_2 e^{2\pi i/3},s_3 e^{4\pi i/3}\right),
    \qquad
    s_2,s_3=\pm1.
\end{gathered}
\end{array}
\label{eq:n_choices_app}
\end{equation}
Charged-like directions are contained in the complex two-component subset as
\begin{equation}
    n_{W^+}=n_{12}^{{\Im},-} \, ,
    \qquad
    n_{W^-}=n_{12}^{{\Im},+} \, .
\end{equation}
In addition to pure $B$ and pure $W$ directions, we include $B$--$W$ gauge-boson superpositions of the form
\begin{equation}
    |u_{BW}\rangle
    =
    \cos\alpha\,|B(\beta_B,\xi_B)\rangle
    +
    e^{i\delta}\sin\alpha\,
    |W(n;\beta_W,\xi_W)\rangle \, ,
\label{eq:BW_superposition_app}
\end{equation}
with
\begin{equation}
    (\beta_B,\xi_B) \, , \,(\beta_W,\xi_W)\in\mathcal H_1 \,,
    \qquad
    n\in\mathcal N_W \, ,
    \qquad
    (\alpha,\delta)\in\mathcal H_2 \, .
\end{equation}
The finite mixing set is
\begin{equation}
    \mathcal H_2
    =
    \left\{
    \left(\frac{\pi}{4},0\right) \, ,
    \left(\frac{\pi}{4},\frac{\pi}{2}\right)
    \right\} \, .
\label{eq:mixing_set_app}
\end{equation}

\subsection{Scalar directions}

In the scalar subspace we define scalar directions by
\begin{equation}
    |\Phi(a)\rangle
    =
    a_0 |H\rangle
    +
    \sum_{I=1}^{3}a_I|\phi^I\rangle \, ,
    \qquad
    |a_0|^2+\sum_{I=1}^{3}|a_I|^2=1 \, .
\label{eq:scalar_direction_app}
\end{equation}
Here $a=(a_0,a_1,a_2,a_3)$ is a unit vector in the scalar space spanned by $H,\phi^1,\phi^2,\phi^3$. In that case, the elementary scalar directions are defined as
\begin{equation}
    \hat a_0=(1,0,0,0) \, ,
    \qquad
    \hat a_1=(0,1,0,0) \, ,
    \qquad
    \hat a_2=(0,0,1,0) \, ,
    \qquad
    \hat a_3=(0,0,0,1) \, .
\end{equation}
The set of representative scalar vectors $a$ used in the scan is defined as $\mathcal{N}_\Phi$ and contains the following representative directions:
\begin{equation}
\begin{array}{ll}
\text{Pure:}
&
\begin{gathered}
    \hat a_j \, ,
    \qquad
    j=0,1,2,3 \, .
\end{gathered}
\\[3mm]
\text{Real two-component:}
&
\begin{gathered}
    a_{jk}^{\Re,\pm}
    =
    \dfrac{1}{\sqrt2}
    \left(\hat a_j\pm\hat a_k\right) \, ,
    \qquad
    0\le j<k\le 3 \, .
\end{gathered}
\\[3mm]
\text{Complex two-component:}
&
\begin{gathered}
    a_{jk}^{\Im,\pm}
    =
    \dfrac{1}{\sqrt2}
    \left(\hat a_j\pm i\hat a_k\right) \, ,
    \qquad
    0\le j<k\le 3 \, .
\end{gathered}
\\[3mm]
\text{Fully coherent:}
&
\begin{gathered}
    a_{\rm all}^{s_1s_2s_3}
    =
    \dfrac{1}{2}
    \left(
        \hat a_0+s_1\hat a_1+s_2\hat a_2+s_3\hat a_3
    \right),
    \qquad
    s_1,s_2,s_3=\pm1 \, .
\end{gathered}
\end{array}
\label{eq:scalar_choices_app}
\end{equation}
Charged-Goldstone directions are contained in the complex two-component subset as
\begin{equation}
    a_{G^+}=a_{12}^{\Im,-} \, ,
    \qquad
    a_{G^-}=a_{12}^{\Im,+} \, .
\end{equation}

\subsection{$V$--$\Phi$ gauge--scalar directions}

In addition to pure gauge and pure scalar directions, we include pairwise $V$--$\Phi$ gauge--scalar superpositions. For $V=B$, we use
\begin{equation}
    |u_{B\Phi}\rangle
    =
    \cos\alpha\,|B(\beta_B,\xi_B)\rangle
    +
    e^{i\delta}\sin\alpha\,|\Phi(a)\rangle \, ,
\label{eq:Bscalar_direction_app}
\end{equation}
while for $V=W$, we use
\begin{equation}
    |u_{W\Phi}\rangle
    =
    \cos\alpha\,|W(n;\beta_W,\xi_W)\rangle
    +
    e^{i\delta}\sin\alpha\,|\Phi(a)\rangle \, .
\label{eq:Wscalar_direction_app}
\end{equation}
The representative choices are
\begin{equation}
    (\beta_B,\xi_B),\,(\beta_W,\xi_W)\in\mathcal H_1 \, ,
    \qquad
    n\in\mathcal N_W \, ,
    \qquad
    a\in\mathcal N_\Phi \, ,
    \qquad
    (\alpha,\delta)\in\mathcal H_2 \, ,
\end{equation}

\subsection{Additional tensorial direction}

For the tensorial sector, we also inspect the non-factorizable weak-isospin--helicity direction
\begin{equation}
    \begin{aligned}
    |W_x^I\rangle&=\frac{1}{\sqrt{2}}\left(|W_-^I\rangle+|W_+^I\rangle\right) \, ,\qquad|W_y^I\rangle=\frac{i}{\sqrt{2}}\left(|W_-^I\rangle-|W_+^I\rangle\right) \, ,\\[1mm]
    |u_{\mathrm{nf}}\rangle&=\frac{1}{\sqrt{2}}\left(|W_x^1\rangle+|W_y^2\rangle\right)=\frac{1}{2}\left(|W_-^1\rangle+|W_+^1\rangle+i|W_-^2\rangle-i|W_+^2\rangle\right) \, .
    \end{aligned}
\end{equation}
Unlike the directions $|W(n;\beta,\xi)\rangle$ above, this state does not factorize into a common helicity direction and a weak-isospin vector.
\clearpage
\section{Block-structure proof of the sector-by-sector unitarity bounds}
\label{app:block_proof}

In this appendix, we show that the marginalized perturbative unitarity bounds obtained by retaining one operator sector at a time coincide with those obtained when all aQGC Wilson coefficients are allowed to vary.
\smallskip

As discussed in Sec.~\ref{sec:unit}, the tree-level coupled-channel matrices $T^J$ are Hermitian. Therefore, imposing perturbative unitarity by requiring
\begin{equation}
    |\lambda_i(T^J)|\leq 1
    \qquad \forall i,J
\end{equation}
is equivalent to demanding
\begin{equation}
    \|T^J\|\leq 1
    \qquad \forall J ,
    \label{eq:norm_unitarity_appendix}
\end{equation}
where $\|\cdot\|$ is the spectral norm.
\smallskip

Let us denote by $c$ the set of Wilson coefficients that belong to one of the sectors considered, and by $d$ all the remaining coefficients. We write $T^{J,{\rm full}}(c,d)$ for the full coupled-channel matrix, and $T^{J,c}(c)$ for the nonzero isolated-sector matrix obtained by setting $d=0$.

At leading order in the high-energy expansion, the field content of the dimension-eight operators determines the two-particle transitions to which each sector can contribute. Working in the electroweak basis, the two-particle channels can be organized as $BB$, $WW$, $BW$, $B\Phi$, $W\Phi$ and $\Phi\Phi$. The $B^4$ and $W^4$ sectors contribute exclusively to the $BB\to BB$ and $WW\to WW$ blocks, respectively, whereas the $B^2W^2$ sector contributes to $BB\leftrightarrow WW$ and $BW\to BW$. Similarly, the $\Phi^2B^2$, $\Phi^2W^2$ and $\Phi^2BW$ sectors contribute, respectively, to the pairs of blocks $\{BB\leftrightarrow\Phi\Phi,\ B\Phi\to B\Phi\}$, $\{WW\leftrightarrow\Phi\Phi,\ W\Phi\to W\Phi\}$ and $\{BW\leftrightarrow\Phi\Phi,\ B\Phi\leftrightarrow W\Phi\}$. Finally, the purely scalar $\Phi^4$ sector contributes exclusively to the $\Phi\Phi\to\Phi\Phi$ block.

Consequently, the contributions associated with each operator sector can be isolated through fixed projectors onto the corresponding two-particle channel subspaces, independently of the remaining Wilson coefficients. The $c$-dependent blocks of the full matrix can therefore be written as
\begin{equation}
    K_{\alpha,J}(c)=P_{\alpha,J}\,T^{J,{\rm full}}(c,d)\,Q_{\alpha,J},
\end{equation}
where \(P_{\alpha,J}\) and \(Q_{\alpha,J}\) are orthogonal projectors and $\alpha$ labels the different blocks dependent on $c$. The isolated-sector matrix therefore has the schematic direct-sum form
\begin{equation}
    T^{J,c}(c)=\bigoplus_{\alpha\in{\cal D}_{J}}K_{\alpha,J}(c)\bigoplus_{\beta\in{\cal O}_{J}}
    \begin{pmatrix}
        0 & K_{\beta,J}(c)\\
        K_{\beta,J}^{\dagger}(c) & 0
    \end{pmatrix},
    \label{eq:isolated_block_sum_appendix}
\end{equation}
where ${\cal D}_{J}$ and ${\cal O}_{J}$ denote, respectively, the sets of principal and off-diagonal blocks.
\smallskip

The spectral norm of the isolated-sector matrix is therefore
\begin{equation}
    \|T^{J,c}(c)\|=\max_{\alpha,\beta}\left\{\|K_{\alpha,J}(c)\|,\|K_{\beta,J}(c)\|\right\}.
\end{equation}
Here we have used the identity
\begin{equation}
    \left\|
    \begin{pmatrix}
        0 & A\\
        A^\dagger & 0
    \end{pmatrix}
    \right\|=\|A\|,
\end{equation}
valid for a general matrix $A$.
\smallskip

Now we define the full and isolated allowed regions as
\begin{equation}
    {\cal R}_{\rm full}=\left\{(c,d):\,\|T^{J,{\rm full}}(c,d)\|\leq 1\quad \forall J\right\},\qquad{\cal R}_c=\left\{c:\,\|T^{J,c}(c)\|\leq 1\quad \forall J\right\},
\end{equation}
and let $\pi_c$ denote the projection onto the coefficients $c$. We also use the standard result that, for a general matrix $A$ and orthogonal projectors $P,Q$, one has~\cite{Bhatia1997}
\begin{equation}
    \|P A Q\|\leq \|A\|.
\end{equation}
Therefore, if $(c,d)\in{\cal R}_{\rm full}$, then every $c$-dependent block satisfies
\begin{equation}
    \|K_{\alpha,J}(c)\|=\|P_{\alpha,J}\hspace{0.6mm}T^{J,{\rm full}}(c,d)\hspace{0.6mm}Q_{\alpha,J}\|\leq\|T^{J,{\rm full}}(c,d)\|\leq 1.
\end{equation}
Using Eq.~\eqref{eq:isolated_block_sum_appendix}, it follows that $\|T^{J,c}(c)\|\leq 1$ for all $J$. Thus $c\in{\cal R}_c$, and hence
\begin{equation}
    \pi_c({\cal R}_{\rm full})\subseteq {\cal R}_c. \label{incl1}
\end{equation}
Conversely, if $c\in{\cal R}_c$, then setting $d=0$ gives
\begin{equation}
    T^{J,{\rm full}}(c,0)=T^{J,c}(c)\oplus 0.
\end{equation}
Therefore $(c,0)\in{\cal R}_{\rm full}$, and
so
\begin{equation}
    {\cal R}_c\subseteq \pi_c({\cal R}_{\rm full}).
    \label{incl2}
\end{equation}
Combining the two inclusions, Eq.~(\ref{incl1}) and Eq.~(\ref{incl2}), gives
\begin{equation}
    \pi_c({\cal R}_{\rm full})={\cal R}_c.
    \label{eq:general_projection_appendix}
\end{equation}
Consequently, for any coefficient $c_i$ in the sector $c$,
\begin{equation}
    \sup_{(c,d)\in{\cal R}_{\rm full}} |c_i|=\sup_{c\in{\cal R}_c} |c_i|.
    \label{eq:general_bound_appendix}
\end{equation}
\clearpage
\section{Full mixed and tensorial eigenvalues}
\label{app:eigenvalues}

\subsection{Full tensorial sector}
We collect here the independent partial-wave eigenvalues of the full tensorial system used to impose perturbative unitarity when deriving the positivity-improved bounds in the $B^2W^2$ sector. For $J=0$, all charge sectors reproduce the particular eigenvalues of the $W^4$ sector given in Eq.~\ref{w4eigen}, while the $Q=0$ and $Q=1$ matrices also reproduce the particular eigenvalues of the $B^2W^2$ sector provided in Eq.~\ref{BWeigen}. Only the $Q=0$ matrix provides four additional independent eigenvalues,
\begin{equation}
    \lambda_i=\frac{s^2}{192\pi\Lambda^4}\,x_i,
    \qquad i=1,2,3,4,
\end{equation}
where the $x_i$ are the roots of
\begin{equation}
    P_T(x)=x^4-\sigma^T_1x^3+\sigma^T_2x^2-\sigma^T_3x+\sigma^T_4 .
\end{equation}
The polynomial coefficients are
\begin{align}
    \sigma^T_1={}&2(a+b),\\[2mm]
    \sigma^T_2={}&
    a^2+4ab+b^2-|d|^2-2|e|^2-2|f|^2-|h|^2,\\[2mm]
    \sigma^T_3={}&
    2ab(a+b)-2b|d|^2
    -2(a+b)\left(|e|^2+|f|^2\right)-2a|h|^2+4\,\Re\left[d^\ast ef+ehf^\ast\right],\\[2mm]
    \sigma^T_4={}&
    a^2b^2-b^2|d|^2-2ab\left(|e|^2+|f|^2\right)-a^2|h|^2
    +|d|^2|h|^2+\left(|e|^2-|f|^2\right)^2
    \nonumber\\[1mm]
    &+4b\,\Re\left[d^\ast ef\right]
    +4a\,\Re\left[ehf^\ast\right]
    -2\,\Re\left[d^\ast e^2h+dh\left(f^\ast\right)^2\right].
\end{align}
Here
\begin{align}
    a&=12\left(4f^T_8+3f^T_9\vphantom{\tilde f}\right),\\[2mm]
    b&=3\left[12f^T_0+4f^T_1+5\left(f^T_2+f^T_3\vphantom{\tilde f}\right)\right],\\[1mm]
    d&=20\left(4f^T_8+f^T_9+4i\tilde f^T_1\right),\\
    e&=3\sqrt3\left[
        f^T_4+4f^T_5+f^T_7
        +8i\left(\tilde f^T_2-\tilde f^T_3\right)
    \right],\\
    f&=\sqrt3\left[
        -f^T_4+12f^T_5+4f^T_6+3f^T_7
        +8i\left(3\tilde f^T_2+3\tilde f^T_3+\tilde f^T_4\right)
    \right],\\
    h&=44f^T_0+28f^T_1+11f^T_2+3f^T_3
    +16i\left(11\tilde f^T_5+7\tilde f^T_6\right).
\end{align}

\subsection{Full mixed sector}
We now collect the independent partial-wave eigenvalues of the full mixed system used to impose perturbative unitarity when deriving the positivity-improved bounds in the $\Phi^2BW$ sector. For $J=0$, the $Q=0$ and $Q=1$ matrices reproduce the particular eigenvalues of the isolated $\Phi^2BW$ sector provided in Eq.~\eqref{j0hhbw}. In addition, the $Q=0$ matrix gives one additional pair of eigenvalues,
\begin{equation}
    \frac{\pm\,s^2}{128\pi\Lambda^4}
    \sqrt{
    \frac{3}{2}\left(8f^M_0-2f^M_1+f^M_7\vphantom{\tilde f}\right)^2
    +8\left(4f^M_2-f^M_3\vphantom{\tilde f}\right)^2
    +128\left[\left(\tilde f^M_1\right)^2+3\left(\tilde f^M_5\right)^2\right]
    } ,
\end{equation}
whereas the $Q=2$ matrix vanishes.
\smallskip

For $J=1$, all charge sectors reproduce the particular eigenvalues of the isolated $\Phi^2W^2$ sector provided in Eq.~\eqref{j1hhww}. The $Q=0$ and $Q=1$ matrices also provide four extra independent eigenvalues,
\begin{equation}
    \lambda_i=\frac{s^2}{3072\pi\Lambda^4}\,x_i,
    \qquad i=1,2,3,4,
\end{equation}
where the $x_i$ are the roots of
\begin{equation}
    P_M(x)=x^4-\sigma^M_1x^3+\sigma^M_2x^2-\sigma^M_3x+\sigma^M_4 .
\end{equation}
The polynomial coefficients are
\begin{align}
    \sigma^M_1={}&2a+b+c,\\[2mm]
    \sigma^M_2={}&
    a^2+2a(b+c)+bc-|d|^2-2|e|^2-|f|^2-|g|^2-|h|^2,\\[2mm]
    \sigma^M_3={}&
    a^2(b+c)+2abc-(b+c)|d|^2-(2a+b+c)|e|^2-(a+b)|f|^2-(a+c)|g|^2-2a|h|^2\nonumber\\[1mm]
    &+2\,\Re\left[
        de^\ast f^\ast+de^\ast g+ef^\ast h+egh
    \right],\\[2mm]
    \sigma^M_4={}&
    a^2bc-a^2|h|^2-ab|e|^2-ab|f|^2-ac|e|^2-ac|g|^2-bc|d|^2+|d|^2|h|^2+|e|^4+|f|^2|g|^2\nonumber\\[1mm]
    &+2a\,\Re\left[ef^\ast h+egh\right]+2b\,\Re\left[de^\ast f^\ast\right]+2c\,\Re\left[de^\ast g\right]-2|e|^2\,\Re\left[fg\right]-2\,\Re\left[d^\ast e^2h+df^\ast gh\right].
\end{align}
Here
\begin{align}
    a&=32f^M_3,\\[2mm]
    b&=8\left(2f^M_1-f^M_7-8f^M_9\vphantom{\tilde f}\right),\\[2mm]
    c&=8\left(2f^M_1-f^M_7+8f^M_9\vphantom{\tilde f}\right),\\[1mm]
    d&=12\left(4f^M_2-f^M_3+4i\tilde f^M_1\right),\\
    e&=16\sqrt3\left(f^M_5-i\tilde f^M_4\right),\\
    f&=2\sqrt3\left[
        -6f^M_4-3f^M_5-5f^M_8
        -i\left(6\tilde f^M_2-5\tilde f^M_3+3\tilde f^M_4\right)
    \right],\\
    g&=2\sqrt3\left[
        -6f^M_4-3f^M_5+5f^M_8
        +i\left(6\tilde f^M_2+5\tilde f^M_3+3\tilde f^M_4\right)
    \right],\\
    h&=24f^M_0-6f^M_1-7f^M_7
    +8i\left(6\tilde f^M_5+5\tilde f^M_6\right).
\end{align}
\clearpage
\section{Polynomial definitions of the $B^2W^2$ and $\Phi^2 B W$ bounds}
\label{app:polynomials}
In this appendix, we collect the polynomial definitions of the numerical constants $\alpha_4$, $\alpha_5$, $\alpha_6$, $\widetilde{\alpha}_{4}$ and $\widetilde{\alpha}_{23}$ that appear in the $B^2W^2$ sector, and of $\beta_{24}$ that appears in the $\Phi^2 BW$ sector, for the positivity-improved perturbative unitarity bounds entering Tables~\ref{tab:multi2}--\ref{tab:multi3}. The constants are defined as the largest positive real roots of the corresponding polynomials, except for $\beta_{24}$, which is defined as the smallest positive real root. The polynomials are presented below in ascending order of degree.

\smallskip

The numerical constants $\alpha_4\approx 25.433$, $\alpha_5\approx 121.837$ and $\widetilde{\alpha}_4\approx 1963.873$ associated, respectively, with the operators $\mathcal{O}^T_4$, $\mathcal{O}^T_5$ and $\cancel{\mathcal{O}}^T_4$ arise from the quartic polynomials
\begin{align}
    P_4(x)&=38382019368\,x^4- 989189566194\,x^3+342136173607\,x^2-310244051433\,x+ 475823711038,\\[2mm]
    P_5(x)&=x^4+3056\,x^3+8586\,x^2-52155576\,x+479615661,\\[2mm]
    \widetilde{P}_4(x)&=x^4-5076480\,x^2-608154048\,x+5898411235584.
\end{align}

The numerical constant $\alpha_{6}\approx 13.937$ associated with the operator $\mathcal{O}^T_6$ arises from the quintic polynomial
\begin{align}
    P_6(x)&=428996538\,x^5-6077033589\,x^4+1261892383 x^3+1097182761\,x^2+4497536398\,x+1985725564.
\end{align}

The numerical constant $\beta_{24}\approx 66.971$ associated with the operators $\mathcal{O}^M_4$ and $\cancel{\mathcal{O}}^M_2$ arises from the eighth-degree polynomial
\begin{align}
    P_{24}(x)={}&319073\,x^{8}-70030016\,x^{7}+778138112\,x^{6}+595944570880\,x^{5}-32867315740672\,x^{4}+\nonumber\\
    &+272643719692288\,x^{3}- 28674483665502208\,x^{2}+1916997146490437632\,x+461439294733549568.
\end{align}

The numerical constant $\widetilde{\alpha}_{23}\approx 3.179$ associated with the operators $\cancel{\mathcal{O}}^T_2$ and $\cancel{\mathcal{O}}^T_3$ arises from the twentieth-degree polynomial
\begin{align}
    \widetilde{P}_{23}(x)={}&2076294995565215990292\,x^{20}+166891870449985214822652\,x^{19}+\nonumber\\
    &+4897066738010497544914758\,x^{18}+48082191016972491439404084\,x^{17}+\nonumber\\
    &-196604430135062323898646135\,x^{16}-3713274460349333692466635800\,x^{15}+\nonumber\\
    &+14279941332012281974083269352\,x^{14}+32048926073882104546973045232\,x^{13}+\nonumber\\
    &-134909786750470374813742592496\,x^{12}+147629047077198927809310142656\,x^{11}+\nonumber\\
    &-2364670240020344589459769852608\,x^{10}+4605428486227932025683285628416\,x^9+\nonumber\\
    &+32719270239528593453020564947968\,x^8-112026212660685106986193366056960\,x^7+\nonumber\\
    &-18714954275203833483813945344000\,x^6+471389210352980827725429103656960\,x^5+\nonumber\\
    &-477192800044452931517654262349824\,x^4-358041821254577951920318285086720\,x^3+\nonumber\\
    &+877951676783895949441373222993920\,x^2-571281963667623939510661036376064\,x+\nonumber\\
    &+124025350895525307399014382043136\vphantom{x^2}.
\end{align}

\end{document}